%% file: Paper_Stop_mu_tau.tex
\RequirePackage{lineno} 
\documentclass[aps,prl,twocolumn,showpacs,superscriptaddress,groupedaddress]{revtex4}  
\usepackage{graphicx}  
\usepackage{dcolumn}   
\usepackage{bm}        
\usepackage{amssymb}   

\newcommand{\stopa}{$\tilde{t}_{1}$}
\newcommand{\astopa}{$\bar{\tilde{t}_{1}}$}
\newcommand{\neuta}{$\tilde{\chi}_{1}^{0}$}
\newcommand{\chara}{$\tilde{\chi}_{1}^{+}$}

\newcommand{\wino}{$\tilde{W}^{+}$}
\newcommand{\higgsino}{$\tilde{H}^{+}$}
\newcommand{\sneut}{${\tilde{\nu}}$}
\newcommand{\bleptsneut}{$b\ell\tilde{\nu}$}
\newcommand{\bmusneut}{$b\mu\tilde{\nu}$}
\newcommand{\btausneut}{$b\tau\tilde{\nu}$}
\newcommand{\bbmutau}{$b\bar{b}\mu\tau \tilde{\nu} \tilde{\nu}$}
\newcommand{\bbemu}{$b\bar{b}e\mu \tilde{\nu} \tilde{\nu}$}
\newcommand{\bbmutaumet}{$b\bar{b}\mu\tau \Met$}
\newcommand{\bbtautau}{$b\bar{b}\tau\tau \tilde{\nu} \tilde{\nu}$}
\newcommand{\bbtautaudec}{$b\bar{b}\tau\tau(\rightarrow\mu \nu \nu)\tilde{\nu} \tilde{\nu}$}
\newcommand{\bbll}{$b\bar{b}\ell\ell^{'} \tilde{\nu} \tilde{\nu}$}
\newcommand{\myb}{$b$}

\newcommand{\mytau}{$\tau$}
\newcommand{\mytauh}{$\tau_h$}
\newcommand{\dzero}{D0}
\newcommand{\fb}{$\rm fb^{-1}$}

\newcommand{\ppbar}{$p\bar{p}$}
\newcommand{\mstop}{$m_{\tilde{t}_{1}}$}
\newcommand{\hmstop}{${1 \over 2}~m_{\tilde{t}_{1}}$}
\newcommand{\tmstop}{2 $m_{\tilde{t}_{1}}$}
\newcommand{\mchara}{$m_{\tilde{\chi}_{1}^{+}}$}
\newcommand{\msneut}{$m_{\tilde{\nu}}$}
\newcommand{\fera}{${\tilde{f}_{1}}$}
\newcommand{\ferb}{${\tilde{f}_{2}}$}
\newcommand{\chira}{$f_{R}$}
\newcommand{\chirb}{$f_{L}$}
\newcommand{\Met}{\mbox{$E\!\!\!\//_{T}$}}
\newcommand{\deltar}{$\sqrt{(\Delta \phi)^{2}+(\Delta \eta)^{2}}$}
\newcommand{\pt}{$p_T$}
\newcommand{\ptmu}{$p_T^{\mu}$}
\newcommand{\ptjet}{$p_T^{\mathrm{jet}}$}

\newcommand{\pttrack}{$p_T^{\mathrm{trk}}$}
\newcommand{\ettau}{$E_T^{\tau}$}
\newcommand{\nntau}{$NN_{\tau}$}
\newcommand{\taua}{$\tau^{\pm}\rightarrow \pi^{\pm} \nu$}
\newcommand{\taub}{$\tau^{\pm}\rightarrow \pi^{\pm} \pi^{0}\nu$}
\newcommand{\tauc}{$\tau^{\pm}\rightarrow \pi^{\pm} \pi^{\pm}\pi^{\mp} ( \pi^{0})\nu$}
\newcommand{\mytaua}{$\tau_1$}
\newcommand{\mytaub}{$\tau_2$}
\newcommand{\mytauc}{$\tau_3$}
\newcommand{\ztautau}{\mbox{$Z/\gamma^*(\rightarrow \tau^+\tau^-)$}+jets}
\newcommand{\zmumu}{\mbox{$Z/\gamma^*(\rightarrow \mu^+\mu^-)$}+jets}
\newcommand{\ww}{$WW$}
\newcommand{\wz}{$WZ$}
\newcommand{\zz}{$ZZ$}
\newcommand{\ttbar}{$t\bar{t}$}
\newcommand{\wjets}{$W$+jets}
\newcommand{\deltamdef}{\mbox{$\Delta m=m_{\tilde{t}_1}-m_{\tilde{\nu}}$}}
\newcommand{\deltam}{$\Delta m$}
\newcommand{\njets}{$N$(jets)}

\newcommand{\ecms}{$\sqrt{s}$}
\newcommand{\deltarmaxtaujet}{$\Delta {\cal{R}}^{\mathrm{max}} (\tau$,jet)}
\newcommand{\deltarmaxmujet}{$\Delta {\cal{R}}^{\mathrm{max}} (\mu$,jet)}
\newcommand{\massleptjet}{Mass(lept,jet)}
\newcommand{\etaleadjet}{$\eta$(leading jet)}
\newcommand{\deltaphiminjetmet}{$\Delta \phi^{\mathrm{min}} $(jet,\Met)}
\newcommand{\deltaphimutaumet}{$\Delta \phi (\mu+\tau,\Met)$}
\newcommand{\massmutau}{Mass($\mu,\tau$)}
\newcommand{\massmujet}{Mass($\mu$,jet)}
\newcommand{\deltaphimutau}{$\Delta \phi (\mu,\tau)$}
\newcommand{\deltaphileadjetmet}{$\Delta \phi $(leading jet,\Met)}
\newcommand{\deltaphimumet}{$\Delta \phi (\mu,\Met)$}
\newcommand{\deltarminmujet}{$\Delta {\cal{R}}^{\mathrm{min}} (\mu$,jet)}
\newcommand{\deltarmintaujet}{$\Delta {\cal{R}}^{\mathrm{min}} (\tau$,jet)}
\newcommand{\deltaphinleadjetmet}{$\Delta \phi $(next-to-leading jet,\Met) }

\begin{document}
\setpagewiselinenumbers
\modulolinenumbers[5]

\vspace*{0.5cm}

\hspace{5.2in} \mbox{FERMILAB-PUB-12-033-E}

\title{Search for pair production of the scalar top quark in muon+tau final states}
\input author_list.tex       
\date{February 9, 2012}

\begin{abstract}
We present a search for the pair production of scalar top quarks  (\stopa), the lightest supersymmetric partners of the top quarks, in \ppbar~collisions at a center-of-mass energy of 1.96 TeV, using data corresponding to an integrated luminosity of \mbox{7.3 \fb} collected with the \dzero~experiment at the Fermilab Tevatron Collider. Each scalar top quark is assumed to decay into a \myb~quark, a charged lepton, and a scalar neutrino (\sneut). We investigate final states arising from \stopa\astopa$\rightarrow$ \bbmutau~and \stopa\astopa$\rightarrow$ \bbtautau. With no significant excess of events observed above the background expected from the standard model, we set exclusion limits on this production process in the (\mstop,\msneut) plane. 
\end{abstract}

\pacs{14.80.Ly, 12.60.Jv, 13.85.Rm}
\maketitle
Supersymmetry (SUSY) \cite{susy} is a space-time symmetry that associates a bosonic partner with each standard model (SM) fermion, and a fermionic counterpart to each SM boson. The mass eigenstates of the scalar fermions, \fera~and \ferb, are the results of the mixing of the SUSY partners of the chiral states \chira~and \chirb. The mass splitting between \fera~and \ferb~depends on the mass of the corresponding fermion. It has been suggested \cite{stoplight} that the large mass of the top quark ($t$) can induce a large splitting between the two stop-mass eigenstates with the consequence that the lightest scalar top quark \stopa~may be sufficiently light to be produced abundantly at the Fermilab Tevatron Collider. If $R$-parity \cite{rparity} is conserved, scalar top quarks would be produced in pairs in \ppbar~collisions, either through gluon fusion or quark-antiquark annihilation.\par
In the minimal supersymmetric extension of the standard model (MSSM) \cite{mssm} with $R$-parity conserved, squarks ($\tilde{q}$) usually decay directly into $\tilde{q}~\rightarrow~q$ \neuta~where the lightest neutralino \neuta~is the lightest supersymmetric particle (LSP). At the Tevatron, this mode is kinematically disfavored for the lightest scalar top quark, because of the large mass of the top quark. If in addition, \mbox{\mstop~$\leq$ $m_b+$\mchara}, where \chara~is the lightest chargino, the decay channel \stopa~$\rightarrow~b$\chara~is not accessible, and the only two-body decay that would be allowed is the flavor changing decay \stopa$\rightarrow~c$\neuta~\cite{cneuta}. Searches related to this channel have been reported by the ALEPH, DELPHI, L3 and OPAL Collaborations \cite{lepstop}. The CDF \cite{cdfcneuta}  and \dzero~\cite{d0cneuta} Collaborations have searched for scalar top quarks in final states with acoplanar charm-jets and large imbalance in transverse momentum (\Met). Other possible decays are the three-body modes \stopa~$\rightarrow~bW$\neuta, \stopa~$\rightarrow~bH^{+}$\neuta, \stopa~ $\rightarrow~b\ell\tilde{\nu}$, and  \stopa~$\rightarrow~b\tilde{\ell}\nu$, where $H^{+}$ is the charged Higgs boson, and $\tilde{\nu}$ and $\tilde{\ell}$ are the sneutrinos and sleptons, superpartners of the neutrinos and leptons, respectively. A possible four-body decay, \stopa~$\rightarrow~b$\neuta$f\bar{f}^{'}$, where $f$ represents a fermion, mediated by virtual top quark, chargino, sbottom, slepton and first/second generation squark exchange,  has also been suggested \cite{stopfourbody}. Searches related to this channel have been reported by the \dzero~Collaboration \cite{dzerofourbody}. It is thought that the three-body decay modes may be important and even dominate the loop-induced $c$\neuta~mode \cite{djouadi}. Searches for scalar top quark pair production in \bbll~final states have been reported by the ALEPH, L3, and OPAL Collaborations \cite{lepstop}. \dzero~\cite{dzerofourbody,dzerostop0,dzerostop1,dzerostop2,dzerostop3} and CDF \cite{cdfstop} have searched for scalar top quark pairs in the final states \bbll, with leptons in $e\mu$, $\mu\mu$, or $ee$ channels. No Tevatron searches have yet considered signatures with hadronically decaying $\tau$ leptons although SUSY could well appear at the Tevatron in final states with taus \cite{sugrawg}. \par
In this letter, we search for scalar top quark pair production in events with $\tau$ leptons, assuming that the branching fraction $B($\stopa~$\rightarrow~$\bleptsneut) = 1 and that the sneutrino is either the LSP or decays invisibly. We search for stop pair production through the decay \stopa\astopa~$\rightarrow$~\bbmutau, or \stopa\astopa~$\rightarrow$~\bbtautaudec, in a data sample corresponding to an integrated luminosity of \mbox{7.3 \fb}~at a center-of-mass energy of 1.96 TeV, collected with the \dzero~detector at the Fermilab Tevatron \ppbar~Collider between April 2002 and July 2010. The signal topology consists of one isolated muon, one isolated $\tau$ lepton, and \Met~coming mainly from undetected sneutrinos, and unreconstructed or mismeasured jets. \par
\input{feynman.tex}

The three-body decay \stopa~$\rightarrow$~\bleptsneut~proceeds mainly through a virtual chargino \chara~(Fig. \ref{fig:feynman}). In the MSSM, \chara~is the lightest mass eigenstate of the charged gaugino-higgsino mass matrix that is a mixing of the wino \wino~and the higgsino \higgsino, SUSY partners of the $W$ boson and the charged Higgs boson, respectively. If \chara~is wino-like, the leptonic decay \chara~$\rightarrow~\ell^{+}\tilde{\nu}$ occurs with equal rate to all lepton flavors. If \chara~is higgsino-like, the decay \chara~$\rightarrow~\tau^{+}\tilde{\nu}$ is enhanced, owing to the large Yukawa coupling of the $\tau$ lepton. In that case, the decay \stopa~$\rightarrow~b \tau \tilde{\nu}$ can be dominant. We consider two scenarios in our search that depend on the composition of the chargino. The wino scenario is defined by $B$(\stopa~$\rightarrow$~\bmusneut) = $B$(\stopa~$\rightarrow~$\btausneut) = 1/3. For the higgsino scenario, we choose $B$(\stopa~$\rightarrow$~\bmusneut) = 0.1 and $B$(\stopa~$\rightarrow$~\btausneut) = 0.8, which correspond to the maximal values reached with a scan of the MSSM parameter space using {\sc susy-hit} \cite{susyhit}. In both scenarios, the signal is a combination of the \bbmutau~and \bbtautau~final states.
\par
The \dzero~detector \cite{d0deta,d0detb,d0detc} is designed to optimize detection and identification of particles arising from \ppbar~interactions and comprises dedicated subsystems surrounding the interaction point. The central tracker resides within a liquid-argon/uranium sampling calorimeter and muon detectors. Charged particles are reconstructed using multi-layer silicon detectors and eight double layers of scintillating fibers in a 1.9 T magnetic field produced by a superconducting solenoid. After passing through the calorimeter, muons are identified using 1.9 T toroids and a muon system composed of three layers of drift tubes and scintillation counters. Events are selected for offline analysis through a three-level trigger system. All events contributing to this analysis are required to pass one of a suite of single-muon triggers based on information from the tracking and muon systems. \par
For each event, the best primary vertex ($pv$) is selected from all the possible reconstructed interaction vertices as the one with smallest probability of originating from a minimum-bias interaction \cite{primver}. To ensure efficient reconstruction, the location of the primary vertex along the beam direction is restricted to $|z_{pv}|<~$60 cm, where $z_{pv}$ is the longitudinal position with respect to the center of the detector.\par

Using central track segments pointing to hit patterns in the muon system, muons are identified in the region $|\eta|~<~$1.8, where $\eta$ is the pseudorapidity \cite{pseudo}. Their trajectories are required to have both drift-tube and scintillator hits that match a track in the central tracker. Muons that are not isolated are rejected if the sum of the transverse momenta of tracks inside a cone of radius $\cal{R}~\equiv~$\deltar~=~0.5 around each muon ($\phi$ being the azimuth), divided by the transverse momentum \ptmu~of the muon, is less than 0.15. The sum of the transverse energies in the calorimeter in an annulus of 0.1$~<~\cal{R}~<~$0.4 around the muon, divided by \ptmu, is also required to be less than 0.15. Only muons with \ptmu~$\geq$ 15 GeV are considered in the analysis. A veto on cosmic ray muons is applied using timing information from the muon system. \par

Decays of $\tau~\rightarrow~$ hadrons+$\nu_\tau$ (called $\tau_h$) are identified with a neural network \cite{nntau} using as input variables $(i)$ calorimeter clusters found with a cone algorithm of $\cal{R}$ = 0.3, $(ii)$ energy in an annular cone 0.3$~\leq~\cal{R}~\leq~$0.5, $(iii)$ electromagnetic (EM) calorimeter subclusters, $(iv)$ the multiplicity of tracks with \pt$~>~$1.5 GeV within $\cal{R}~<~$0.5 of the direction of the $\tau$ lepton, and $(v)$ consistency of the invariant mass of the hadron system with that of $\tau$ decay. Three neural networks \nntau~are trained to identify tau decays corresponding to \taua~(\mytaua), \taub~(\mytaub), and \tauc~(\mytauc). 
In addition, a selection on the output of $NN_{el}$, a neural network trained to separate \mytaub~from electrons of similar signatures, is applied. The minimum transverse energy of the $\tau_h$ measured in the calorimeter, \ettau, is 12.5 GeV for \mytaua~and \mytaub~and 15 GeV for \mytauc. The sum of the transverse momenta of the \mytau-associated tracks, \pttrack, is required to exceed 7, 5, and 10 GeV, for \mytaua, \mytaub, and \mytauc, respectively. In addition, at least one track with \mbox{\pt $~>~$ 7 GeV} is required for \mytauc. Finally, only candidates with $|\eta_{\tau_{h}}|~<~$1 and \pttrack/\ettau $>$ (0.65, 0.5, 0.5) for (\mytaua, \mytaub, \mytauc) are selected. \par

Jets are reconstructed from energies deposited in calorimeter towers using an iterative midpoint cone algorithm \cite{conealgo}, with a cone radius \mbox{${\cal R }$ = 0.5}.
 Jet energies are calibrated to the particle-level jets using correction factors derived primarily from the transverse momentum balance in photon plus jets events \cite{jes}. Only jets with \ptjet~$>$ 15 GeV and $|\eta|~<~$2.5 are considered in this analysis and  jets in the vicinity of a \mytauh~candidate ($\cal{R}~<$ 0.5) are discarded. \par
The \Met~is calculated from the calorimeter energy, corrected for jet, EM, and \mytau~energy scales and for the transverse momentum of selected muons.\par

\input{data_mc_presel.tex}
Monte Carlo (MC) events for signal are simulated using {\sc madgraph/madevent} \cite{madgraph} and {\sc pythia} \cite{pythia} for parton-level generation and hadronization, respectively. We consider a range of scalar top quark mass values from 100 to 200 GeV, generated in steps of 20 GeV. The range of probed sneutrino masses extends from 40 to 140 GeV in steps of 20 GeV. For each hypothesis, the MSSM parameters are estimated from {\sc suspect} \cite{suspect} and {\sc sdecay} \cite{sdecay}. The next-to-leading order (NLO) scalar top quark pair production cross section is calculated with {\sc prospino 2.0} \cite{prospino}, using  {\small CTEQ6.1M} parton distribution functions (PDF)~\cite{pumplin,stump}. The calculations are performed with the renormalization and factorization scales $\mu_{r,f}$ equal to the scalar top quark mass \mstop, \hmstop, and \tmstop~to estimate uncertainty on the nominal value through the impact of the two excursions. These uncertainties are combined quadratically with uncertainties on the PDF ~\cite{pumplin,stump} to provide a total theoretical uncertainty of 18\% to 20\% on the scalar top quark cross section. \par
The kinematics for signal are determined both by \mstop~and by the mass difference \deltamdef. The \pt~of the leptons and $b$ quarks decrease on average for smaller values of \deltam,  and \Met~is correlated with both \mstop~and \deltam. We choose two signal points [\mstop,\msneut] = (180,60) GeV and (120,80) GeV, labeled ``Signal A'' and ``Signal B'' in the following, to illustrate the impact of the selection criteria for large \mstop~and \deltam~(Signal A) and for low \mstop~and \deltam~(Signal B).

The dominant SM backgrounds to the pair production of scalar top quarks are  from \ztautau; \zmumu; diboson production (\ww, \wz, \zz); \ttbar; \wjets, and instrumental background from multijet (MJ) processes. All but the latter are estimated through MC simulations. Vector boson pair production is simulated with {\sc pythia}, while the other backgrounds are simulated at the parton level using {\sc alpgen} \cite{alpgen} and  {\sc pythia} for hadronization and parton showering. \par
Correction factors for MC estimated from data are applied to lessen the impact of minor mismodeling of detector response. These corrections are related to the instantaneous luminosity, the position of the beam spot, identification efficiencies for $\mu$ and $\tau$, vector boson \pt, and jet, muon, and \mytauh~energy resolutions.\par

The instrumental background originates either from incorrectly-identified, isolated muons (arising for example from semi-leptonic $b$ decays) or from misidentified \mytauh~ (jets mimicking \mytauh~signatures). This is estimated
~by changing the requirements on muon isolation and on the \mytauh~\nntau~outputs for each type of tau after the subtraction of the MC contributions corresponding to non-instrumental background. Normalization factors for these samples are estimated assuming that MJ processes have equal amounts of like-charge and opposite-charge $\mu\tau$ events. 

The search of scalar top quark pairs proceeds in three steps: two event selections, labeled ``Selection-1'' and ``Selection-2'' below, and then a multivariate analysis. \par
Selection-1 requires candidates to contain exactly one muon and \mytauh~of opposite electric charge, and to have a minimum separation of $\cal{R}(\mu$,\mytauh)$~>~$ 0.5 between the two leptons. No specific requirement on jets is applied at this stage, but events having a jet in the vicinity of the muon ($\Delta \cal{R}~<~$0.5) are rejected. Events with low values of the azimuthal angle difference between the leptons and \Met, which are often due to issues with lepton reconstruction, are  removed by requiring $\Delta \phi(\mu,\Met)~>~$0.5 and $\Delta \phi(\tau_{h},\Met)~>~$0.5. At this stage of the analysis, 3387 data events remain while \mbox{3453 $\pm$ 29 (stat) $\pm$ 440 (syst)} events are expected from background. The main background is from \ztautau, \wjets, and MJ events, as can be seen in Table \ref{tab:presel}. Signal efficiencies do not exceed 4\% for large \deltam, and are lower than 0.1\% for \mbox{\deltam$~<~$ 20 GeV}. \par

\input{plots_presel.tex}
Jets in scalar top quark pair events originate mainly from the hadronization of $b$ quarks, whereas in \ztautau~and \wjets~backgrounds jets correspond predominantly to initial-state gluon radiation that provides a lower jet multiplicity. To maintain sensitivity to low \deltam~signals, while rejecting a substantial part of the background, at least one jet is required in each event, which corresponds to Selection-2. Fig.\ \ref{fig:presel} shows the jet multiplicity for the different \mytauh~decays. After this selection, 893 events remain, while a total background of 905 $\pm$ 10 (stat) $\pm$ 127 (syst) events is expected (see Table \ref{tab:selection1}).\par

\input{data_mc_selection1.tex}
The separation of scalar top quark signal and background is improved through an implementation of Boosted Decision Trees (BDT) \cite{bdt}. A decision tree classifies events on the basis of cumulative selection criteria that define disjoint subsets of events, each with a different signal purity. The decision tree is redefined iteratively by creating subsets of events called nodes. Each node is split into two subsets on the basis of the strongest discriminant for that sample. An impurity measure $i$ is estimated for each node from the weighted number of signal $S$ and background $B$ events in the node. For a given split, the decrease of impurity $\Delta i~=~i(S,B)-i(S_L,B_L)-i(S_R,B_R)$, where $L$ and $R$ stand for left and right daughter nodes, is calculated. The best splitting gives the largest $\Delta i$. We use the Gini index \cite{bdt} defined as $SB/(S+B)^2$ as a measure of impurity. Terminal nodes are called leaves. Each leaf has a purity value defined by $S/(S+B)$. One of the main advantage of decision trees over analyses using simple requirements is that events that fail any individual selection criteria continue to be considered by the algorithm. \par
\input{bdtvar.tex}
The performance of the decision tree is improved by the boosting technique \cite{boosting}. The basic principle is to create a tree, calculate an associated uncertainty, and create a new tree with a smaller uncertainty by re-weighting the misclassified events. We use adaptative boosting, known in the literature as AdaBoost \cite{boosting}. The associated uncertainty $\epsilon_n$ of a tree indexed by $n$ is estimated as the fraction of misclassified events. The boosting weight of the $n^{th}$ tree is $\alpha_n~=~\beta~\ln((1-\epsilon_n)/\epsilon_n)$ where $\beta$ is an empirically determined parameter called boosting parameter. Misclassified events are given an additional multiplicative weight of $e^{\alpha_n}$ and the resulting new tree indexed by $n+1$ is used to retrain the BDT and reduce the number of misclassified events. This procedure is repeated $N$ times, where $N$ is the number of boosting cycles. For the scalar top quark search, the best signal to background separation occurs for $\beta~=~0.5$ and $N~=~40$. \par 
\input{plots_bdt.tex}
To optimize the sensitivity of the analysis, three subsamples are selected according to the jet multiplicity per event: \njets = 1, \njets = 2, \njets $>$ 2. Since there are many SUSY mass points, and their characteristics differ significantly, the generated BDTs are trained and tested for each sample and each SUSY point using the implementation of the {\sc tmva} \cite{tmva} library. The five most sensitive input variables for the BDTs trained with samples of Signal A and Signal B are given in Table \ref{tab:bdtvars}. To minimize bias, samples are split into three parts: 1/3 for training, 1/3 for testing  and 1/3 for analysis. The distributions of the BDT outputs trained with Signals A and B are shown in Figs.\ \ref{fig:bdts_wino} and \ref{fig:bdts_higgsino} for the wino and higgsino scenarios, respectively. \par
The predicted numbers of background and signal events depend on measurements and parametrizations that have non-negligible systematic uncertainties, which can either affect exclusively the normalizations of backgrounds or the signal efficiency, or modify also the differential distribution of the BDT discriminant. The main sources involve muon identification and reconstruction efficiencies (2\%), \mytauh~identification and reconstruction (10\%, 4\%, and 5\% for \mytaua, \mytaub~and \mytauc, respectively), trigger (5\%), luminosity (6.1\%) \cite{d0lumi}, jet energy calibration (3.2\% for the background, 1.5\% to 2.6\% for signal), jet identification efficiency and energy resolution (5\% for the background,  1\% to 6\% for signal). Systematic uncertainties related to reconstructed objects are estimated by changing each quantity by one standard deviation (s.d), and gauging the impact on the final measurement.  Additional uncertainties arise from the choice of PDF, which affects the cross sections for the background components (6.3\% for \ztautau~and \zmumu, 15\% for \wjets~and 10\% for \ttbar; 5.6\%, 8.1\%, and 5.5\% for \ww, \wz, and \zz, respectively) and for signal (18\% to 20\%). To estimate the systematic uncertainties related to instrumental background, a MC scalar top quark signal is added as a background contribution during the stages that consider distribution and normalization of the MJ background. Each scalar top quark signal is considered, and the largest relative changes in the distribution and the normalization of the instrumental background are thereby estimated. The resulting uncertainties correspond to 5\% and 10\% for the normalization and the functional dependence, respectively. \par
\input{exclusion_nosigma.tex}
There is no significant excess of events observed above the predicted background, and we therefore combine the numbers of expected signal and background events, with their corresponding uncertainties, and the numbers of events observed in data obtained from the BDT outputs for each SUSY point, to calculate upper limits on the cross sections for signal  at the 95\% CL using the modified frequentist approach \cite{cls}. The bins of the BDT outputs corresponding to \njets = 1, 2, and $>$ 2, are treated as separate channels, and their likelihoods are combined taking into account the correlations of both systematic uncertainties affecting exclusively the normalization of backgrounds and signal efficiencies and also of those that change the distribution of the BDT discriminant. The limits are calculated using the confidence level $CL_S=CL_{S+B}/CL_B$ where $CL_{S+B}$ and $CL_B$ are the confidence levels for the signal+background and background-only hypotheses, respectively \cite{cls}. Exclusion regions are given on Figs.\ \ref{fig:limite_wino}  and \ref{fig:limite_higgsino} as a function of the scalar top quark and sneutrino masses. These results are obtained under the assumption $B$(\stopa~$\rightarrow$~\bmusneut) = $B$(\stopa~$\rightarrow$~\btausneut) = 1/3 (Fig.\ \ref{fig:limite_wino}, wino scenario) and $B$(\stopa~$\rightarrow$~\bmusneut) = 0.1, $B$(\stopa~$\rightarrow~$\btausneut) = 0.8 (Fig.\ \ref{fig:limite_higgsino}, higgsino scenario). For larger mass differences between the scalar top quark and the sneutrino, a scalar top quark mass lower than 200 GeV is excluded. The search is sensitive to a possible signal in the mass region
up to \deltam~=~60 GeV for \mstop~=~140 GeV, with  the observed limit being within one standard deviation of the expected limit. \par

In summary, a search for scalar top quark pair production in \ppbar~collisions at \ecms~=~1.96 TeV has been performed in a dataset corresponding to an integrated luminosity of \mbox{7.3 \fb}. Events containing one muon, one $\tau$ decaying hadronically, at least one jet, and missing transverse energy have been considered in this analysis. No evidence is found for the production of the lightest scalar top quark, and 95\% CL exclusion limits are set in the plane [\mstop,\msneut]. The largest scalar top quark mass excluded is 200 GeV for a sneutrino mass of 45 GeV, and the largest sneutrino mass excluded is 85 GeV for a scalar top quark mass of 160 GeV. This is the first Tevatron limit obtained from a study of final states containing \mytau~leptons from \stopa\astopa$\rightarrow$ \bbmutaumet.

\input{acknowledgement.tex}   

\end{document}

%% file: author_list.tex
%
\affiliation{Universidad de Buenos Aires, Buenos Aires, Argentina}
\affiliation{LAFEX, Centro Brasileiro de Pesquisas F{\'\i}sicas, Rio de Janeiro, Brazil}
\affiliation{Universidade do Estado do Rio de Janeiro, Rio de Janeiro, Brazil}
\affiliation{Universidade Federal do ABC, Santo Andr\'e, Brazil}
\affiliation{Instituto de F\'{\i}sica Te\'orica, Universidade Estadual Paulista, S\~ao Paulo, Brazil}
\affiliation{University of Science and Technology of China, Hefei, People's Republic of China}
\affiliation{Universidad de los Andes, Bogot\'{a}, Colombia}
\affiliation{Charles University, Faculty of Mathematics and Physics, Center for Particle Physics, Prague, Czech Republic}
\affiliation{Czech Technical University in Prague, Prague, Czech Republic}
\affiliation{Center for Particle Physics, Institute of Physics, Academy of Sciences of the Czech Republic, Prague, Czech Republic}
\affiliation{Universidad San Francisco de Quito, Quito, Ecuador}
\affiliation{LPC, Universit\'e Blaise Pascal, CNRS/IN2P3, Clermont, France}
\affiliation{LPSC, Universit\'e Joseph Fourier Grenoble 1, CNRS/IN2P3, Institut National Polytechnique de Grenoble, Grenoble, France}
\affiliation{CPPM, Aix-Marseille Universit\'e, CNRS/IN2P3, Marseille, France}
\affiliation{LAL, Universit\'e Paris-Sud, CNRS/IN2P3, Orsay, France}
\affiliation{LPNHE, Universit\'es Paris VI and VII, CNRS/IN2P3, Paris, France}
\affiliation{CEA, Irfu, SPP, Saclay, France}
\affiliation{IPHC, Universit\'e de Strasbourg, CNRS/IN2P3, Strasbourg, France}
\affiliation{IPNL, Universit\'e Lyon 1, CNRS/IN2P3, Villeurbanne, France and Universit\'e de Lyon, Lyon, France}
\affiliation{III. Physikalisches Institut A, RWTH Aachen University, Aachen, Germany}
\affiliation{Physikalisches Institut, Universit{\"a}t Freiburg, Freiburg, Germany}
\affiliation{II. Physikalisches Institut, Georg-August-Universit{\"a}t G\"ottingen, G\"ottingen, Germany}
\affiliation{Institut f{\"u}r Physik, Universit{\"a}t Mainz, Mainz, Germany}
\affiliation{Ludwig-Maximilians-Universit{\"a}t M{\"u}nchen, M{\"u}nchen, Germany}
\affiliation{Fachbereich Physik, Bergische Universit{\"a}t Wuppertal, Wuppertal, Germany}
\affiliation{Panjab University, Chandigarh, India}
\affiliation{Delhi University, Delhi, India}
\affiliation{Tata Institute of Fundamental Research, Mumbai, India}
\affiliation{University College Dublin, Dublin, Ireland}
\affiliation{Korea Detector Laboratory, Korea University, Seoul, Korea}
\affiliation{CINVESTAV, Mexico City, Mexico}
\affiliation{Nikhef, Science Park, Amsterdam, the Netherlands}
\affiliation{Radboud University Nijmegen, Nijmegen, the Netherlands and Nikhef, Science Park, Amsterdam, the Netherlands}
\affiliation{Joint Institute for Nuclear Research, Dubna, Russia}
\affiliation{Institute for Theoretical and Experimental Physics, Moscow, Russia}
\affiliation{Moscow State University, Moscow, Russia}
\affiliation{Institute for High Energy Physics, Protvino, Russia}
\affiliation{Petersburg Nuclear Physics Institute, St. Petersburg, Russia}
\affiliation{Instituci\'{o} Catalana de Recerca i Estudis Avan\c{c}ats (ICREA) and Institut de F\'{i}sica d'Altes Energies (IFAE), Barcelona, Spain}
\affiliation{Stockholm University, Stockholm and Uppsala University, Uppsala, Sweden}
\affiliation{Lancaster University, Lancaster LA1 4YB, United Kingdom}
\affiliation{Imperial College London, London SW7 2AZ, United Kingdom}
\affiliation{The University of Manchester, Manchester M13 9PL, United Kingdom}
\affiliation{University of Arizona, Tucson, Arizona 85721, USA}
\affiliation{University of California Riverside, Riverside, California 92521, USA}
\affiliation{Florida State University, Tallahassee, Florida 32306, USA}
\affiliation{Fermi National Accelerator Laboratory, Batavia, Illinois 60510, USA}
\affiliation{University of Illinois at Chicago, Chicago, Illinois 60607, USA}
\affiliation{Northern Illinois University, DeKalb, Illinois 60115, USA}
\affiliation{Northwestern University, Evanston, Illinois 60208, USA}
\affiliation{Indiana University, Bloomington, Indiana 47405, USA}
\affiliation{Purdue University Calumet, Hammond, Indiana 46323, USA}
\affiliation{University of Notre Dame, Notre Dame, Indiana 46556, USA}
\affiliation{Iowa State University, Ames, Iowa 50011, USA}
\affiliation{University of Kansas, Lawrence, Kansas 66045, USA}
\affiliation{Kansas State University, Manhattan, Kansas 66506, USA}
\affiliation{Louisiana Tech University, Ruston, Louisiana 71272, USA}
\affiliation{Boston University, Boston, Massachusetts 02215, USA}
\affiliation{Northeastern University, Boston, Massachusetts 02115, USA}
\affiliation{University of Michigan, Ann Arbor, Michigan 48109, USA}
\affiliation{Michigan State University, East Lansing, Michigan 48824, USA}
\affiliation{University of Mississippi, University, Mississippi 38677, USA}
\affiliation{University of Nebraska, Lincoln, Nebraska 68588, USA}
\affiliation{Rutgers University, Piscataway, New Jersey 08855, USA}
\affiliation{Princeton University, Princeton, New Jersey 08544, USA}
\affiliation{State University of New York, Buffalo, New York 14260, USA}
\affiliation{Columbia University, New York, New York 10027, USA}
\affiliation{University of Rochester, Rochester, New York 14627, USA}
\affiliation{State University of New York, Stony Brook, New York 11794, USA}
\affiliation{Brookhaven National Laboratory, Upton, New York 11973, USA}
\affiliation{Langston University, Langston, Oklahoma 73050, USA}
\affiliation{University of Oklahoma, Norman, Oklahoma 73019, USA}
\affiliation{Oklahoma State University, Stillwater, Oklahoma 74078, USA}
\affiliation{Brown University, Providence, Rhode Island 02912, USA}
\affiliation{University of Texas, Arlington, Texas 76019, USA}
\affiliation{Southern Methodist University, Dallas, Texas 75275, USA}
\affiliation{Rice University, Houston, Texas 77005, USA}
\affiliation{University of Virginia, Charlottesville, Virginia 22901, USA}
\affiliation{University of Washington, Seattle, Washington 98195, USA}
\author{V.M.~Abazov} \affiliation{Joint Institute for Nuclear Research, Dubna, Russia}
\author{B.~Abbott} \affiliation{University of Oklahoma, Norman, Oklahoma 73019, USA}
\author{B.S.~Acharya} \affiliation{Tata Institute of Fundamental Research, Mumbai, India}
\author{M.~Adams} \affiliation{University of Illinois at Chicago, Chicago, Illinois 60607, USA}
\author{T.~Adams} \affiliation{Florida State University, Tallahassee, Florida 32306, USA}
\author{G.D.~Alexeev} \affiliation{Joint Institute for Nuclear Research, Dubna, Russia}
\author{G.~Alkhazov} \affiliation{Petersburg Nuclear Physics Institute, St. Petersburg, Russia}
\author{A.~Alton$^{a}$} \affiliation{University of Michigan, Ann Arbor, Michigan 48109, USA}
\author{G.~Alverson} \affiliation{Northeastern University, Boston, Massachusetts 02115, USA}
\author{M.~Aoki} \affiliation{Fermi National Accelerator Laboratory, Batavia, Illinois 60510, USA}
% ...
%
%
\collaboration{The D0 Collaboration\footnote{with visitors from
$^{a}$Augustana College, Sioux Falls, SD, USA,
$^{b}$The University of Liverpool, Liverpool, UK,
$^{c}$UPIITA-IPN, Mexico City, Mexico,
$^{d}$SLAC, Menlo Park, CA, USA,
$^{e}$University College London, London, UK,
$^{f}$Centro de Investigacion en Computacion - IPN, Mexico City, Mexico,
$^{g}$ECFM, Universidad Autonoma de Sinaloa, Culiac\'an, Mexico,
and 
$^{h}$Universit{\"a}t Bern, Bern, Switzerland.
$^{i}$DESY, Hamburg, Germany,
$^{\ddag}$Deceased.
}} \noaffiliation
\vskip 0.25cm

%% file: feynman.tex
\begin{figure}[!htpb]
\includegraphics[width=5.5cm]{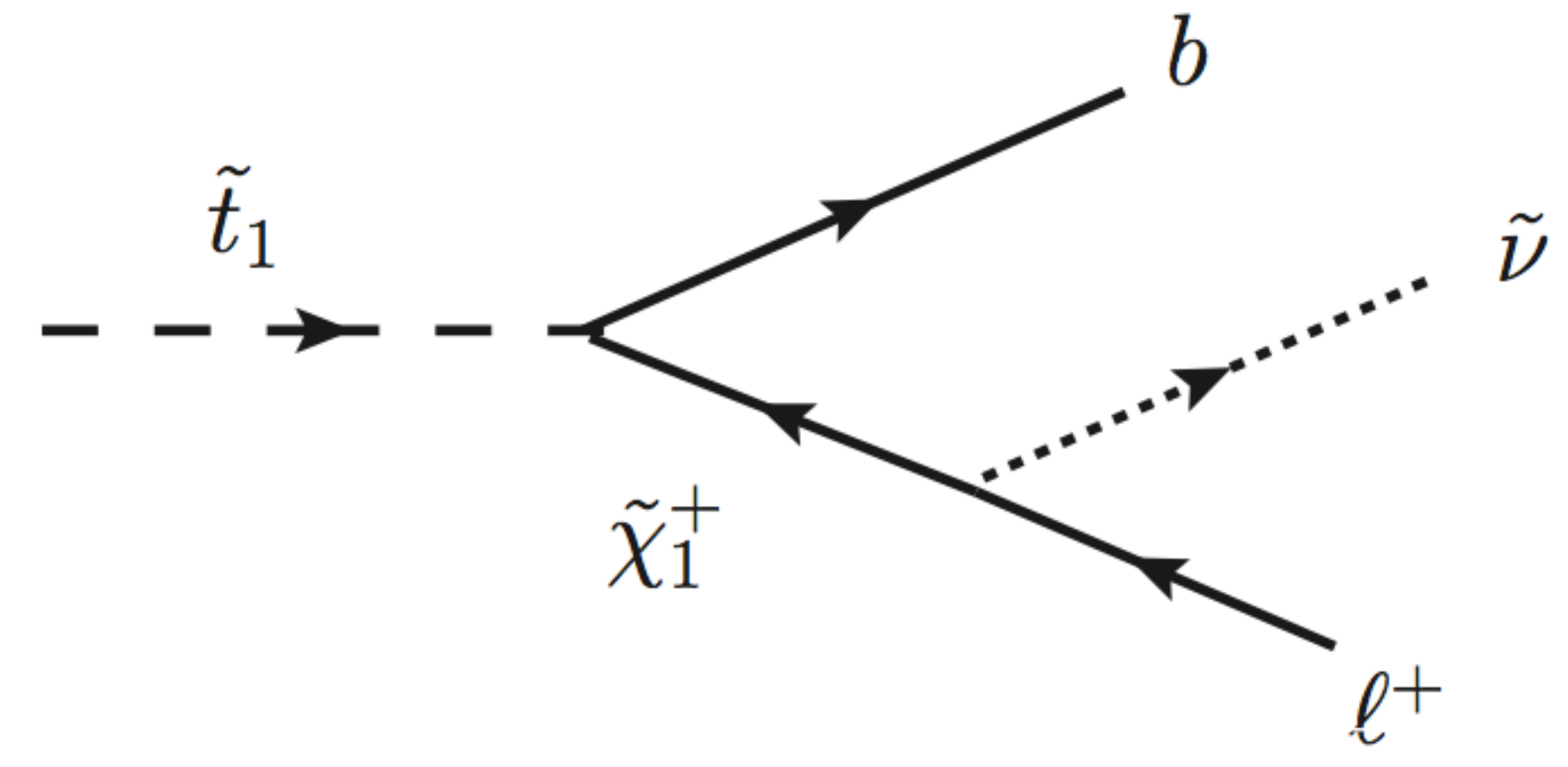}
\caption{Diagram contributing to the three-body decay \mbox{\stopa $\rightarrow$ \bleptsneut}.}\label{fig:feynman}
\end{figure}

%% file: data_mc_presel.tex
\begin{table*}[!htpb]
 \caption{Numbers of events observed and expected from SM background
   processes and the two signal samples A and B at the Selection-1
   stage for \mytauh~(\mytaua, \mytaub, \mytauc) and their sum. The uncertainties quoted
   in the Table are statistical.}\label{tab:presel}
 \begin{tabular}{cD{,}{\,\pm\,}{-1}D{,}{\,\pm\,}{-1}D{,}{\,\pm\,}{-1}D{,}{\,\pm\,}{-1}} 
\hline
\hline
\multicolumn{1}{c}{Process} & \multicolumn{1}{c}{\mytaua}&
\multicolumn{1}{c}{\mytaub} & \multicolumn{1}{c}{\mytauc}
& \multicolumn{1}{c}{all \mytau} \\
\hline
 \multicolumn{1}{c}{\ztautau} &  \multicolumn{1}{c}{162.6}    &  \multicolumn{1}{c}{994.4}   &  \multicolumn{1}{c}{352.3}     &  \multicolumn{1}{c}{1509.4} \\  
 \multicolumn{1}{c}{\zmumu} &  \multicolumn{1}{c}{38.7}     & \multicolumn{1}{c}{91.6}       &  \multicolumn{1}{c}{48.3}       &  \multicolumn{1}{c}{178.6} \\
 \multicolumn{1}{c}{diboson } &  \multicolumn{1}{c}{7.5}       &  \multicolumn{1}{c}{40.5}        &  \multicolumn{1}{c}{16.9}    &  \multicolumn{1}{c}{65.0}\\ 
 \multicolumn{1}{c}{\ttbar} &  \multicolumn{1}{c}{3.2}            &  \multicolumn{1}{c}{27.3}        &  \multicolumn{1}{c}{10.7}     &  \multicolumn{1}{c}{41.3} \\
 \multicolumn{1}{c}{\wjets} &  \multicolumn{1}{c}{125.1}        &  \multicolumn{1}{c}{631.1}     &  \multicolumn{1}{c}{421.1}   &  \multicolumn{1}{c}{1177.2} \\
 \multicolumn{1}{c}{Instrumental} &  \multicolumn{1}{c}{54.7} &  \multicolumn{1}{c}{233.1}     &  \multicolumn{1}{c}{193.4}   &  \multicolumn{1}{c}{481.3} \\
\hline
 \multicolumn{1}{c}{Background total} & 392.0,6.0 & 2018.1,16.9 & 1042.8,12.6 & 3452.8,28.6 \\
 \multicolumn{1}{c}{Data} & \multicolumn{1}{c}{388} & \multicolumn{1}{c}{1937} & \multicolumn{1}{c}{1062} & \multicolumn{1}{c}{3387} \\
\hline
 \multicolumn{1}{c}{Wino scenario}       &  &  &  & \\
\multicolumn{1}{c}{Signal A }       & 2.3,0.3  & 17.4,0.8   & 4.7,0.4  & 24.4,1.0 \\
 \multicolumn{1}{c}{Signal B}       &  4.4,1.3        & 21.4,3.0        & 7.1,1.7        & 32.9,3.7 \\
\hline
 \multicolumn{1}{c}{Higgsino scenario}       & &   & & \\
 \multicolumn{1}{c}{Signal A} & 2.9,0.3  & 20.5,0.9  & 5.3,0.5  & 28.6,1.1 \\
 \multicolumn{1}{c}{Signal B } &  3.8,1.2        & 20.5,2.9        &
 6.9,1.6  & 31.1,3.6 \\
\hline
\hline
  \end{tabular}
 \end{table*}

%% file: plots_presel.tex
\begin{figure*}[htbp]
\includegraphics[scale=0.3]{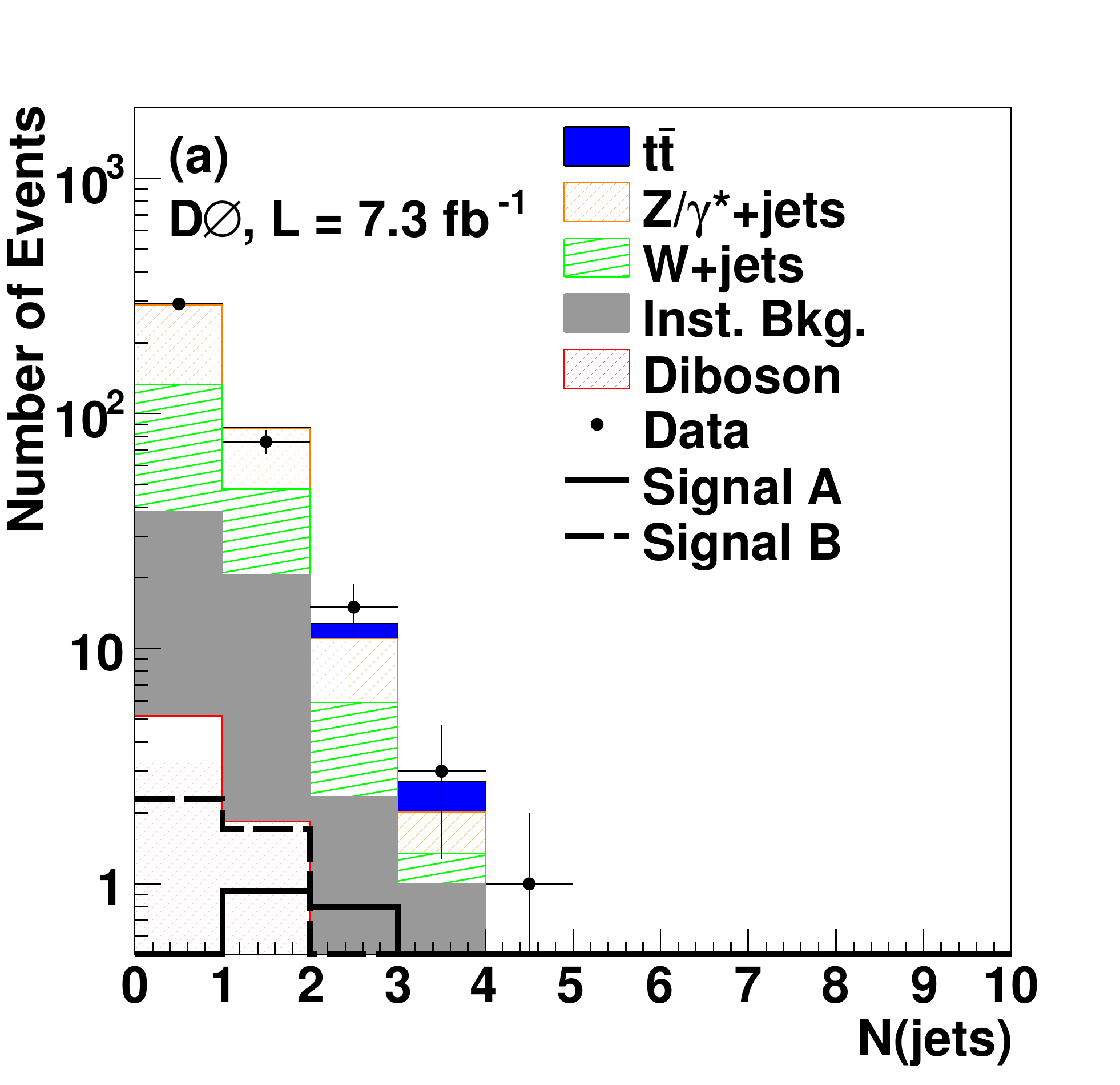}
\includegraphics[scale=0.3]{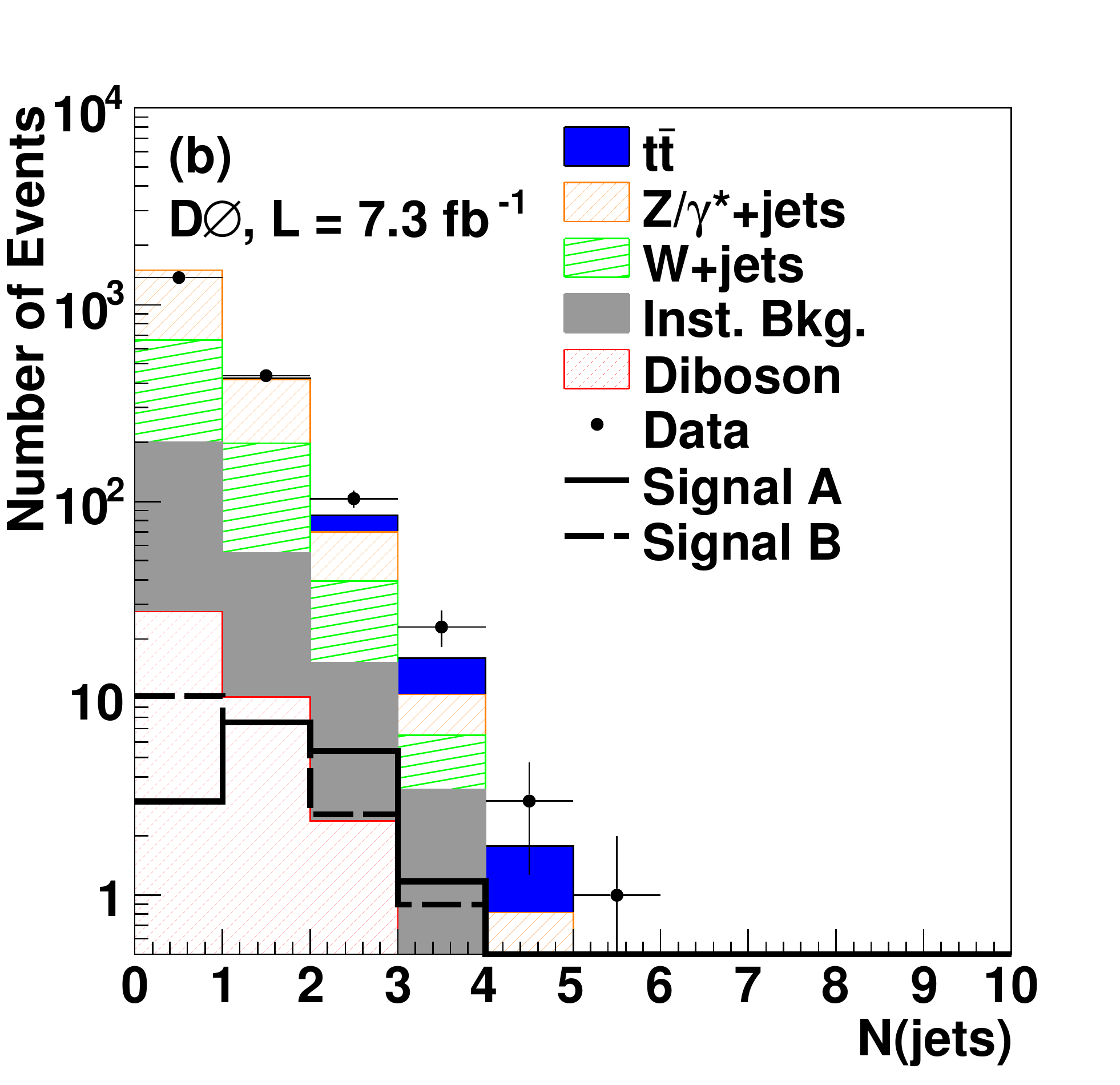} \\
\includegraphics[scale=0.3]{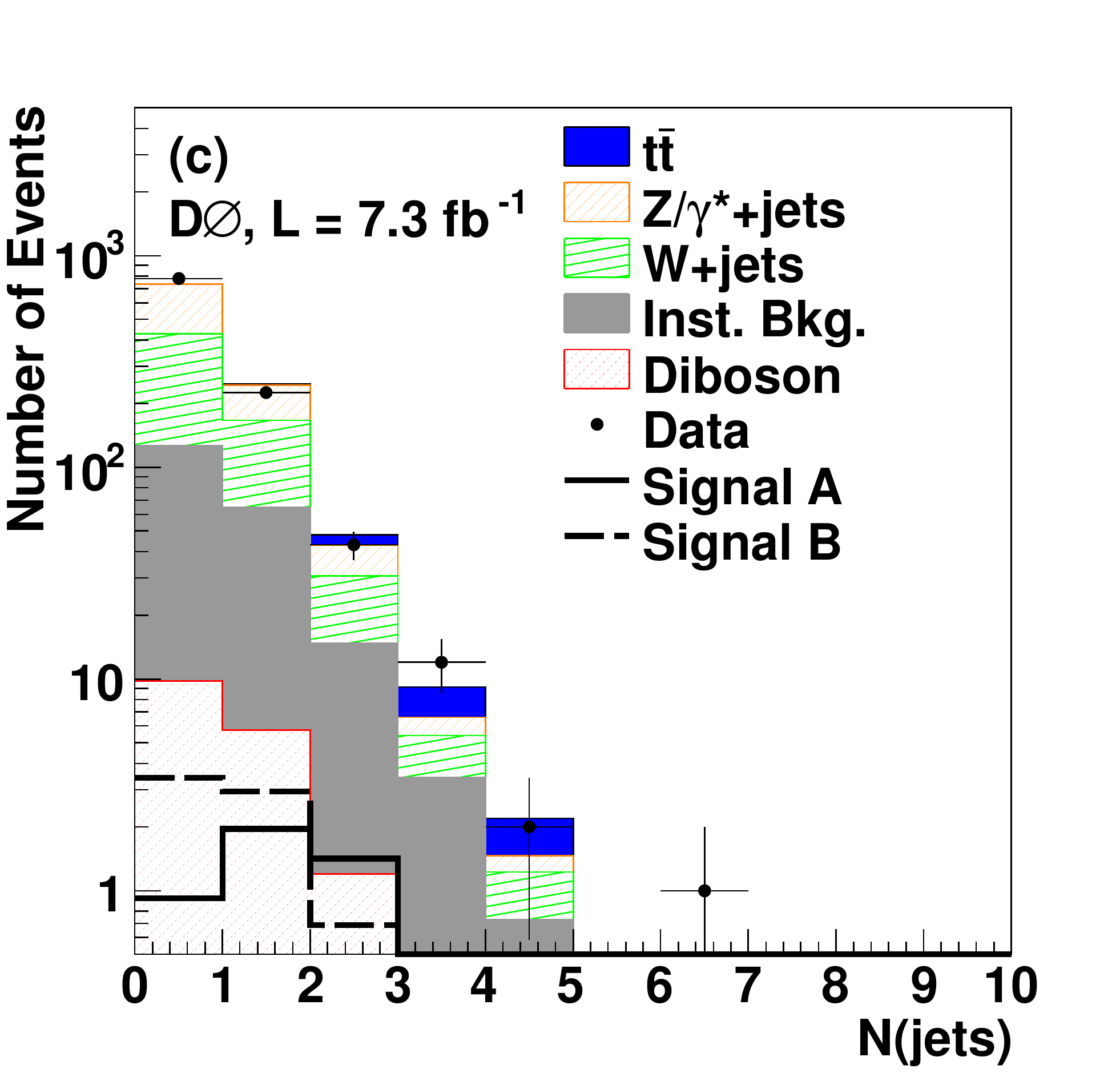}
\includegraphics[scale=0.3]{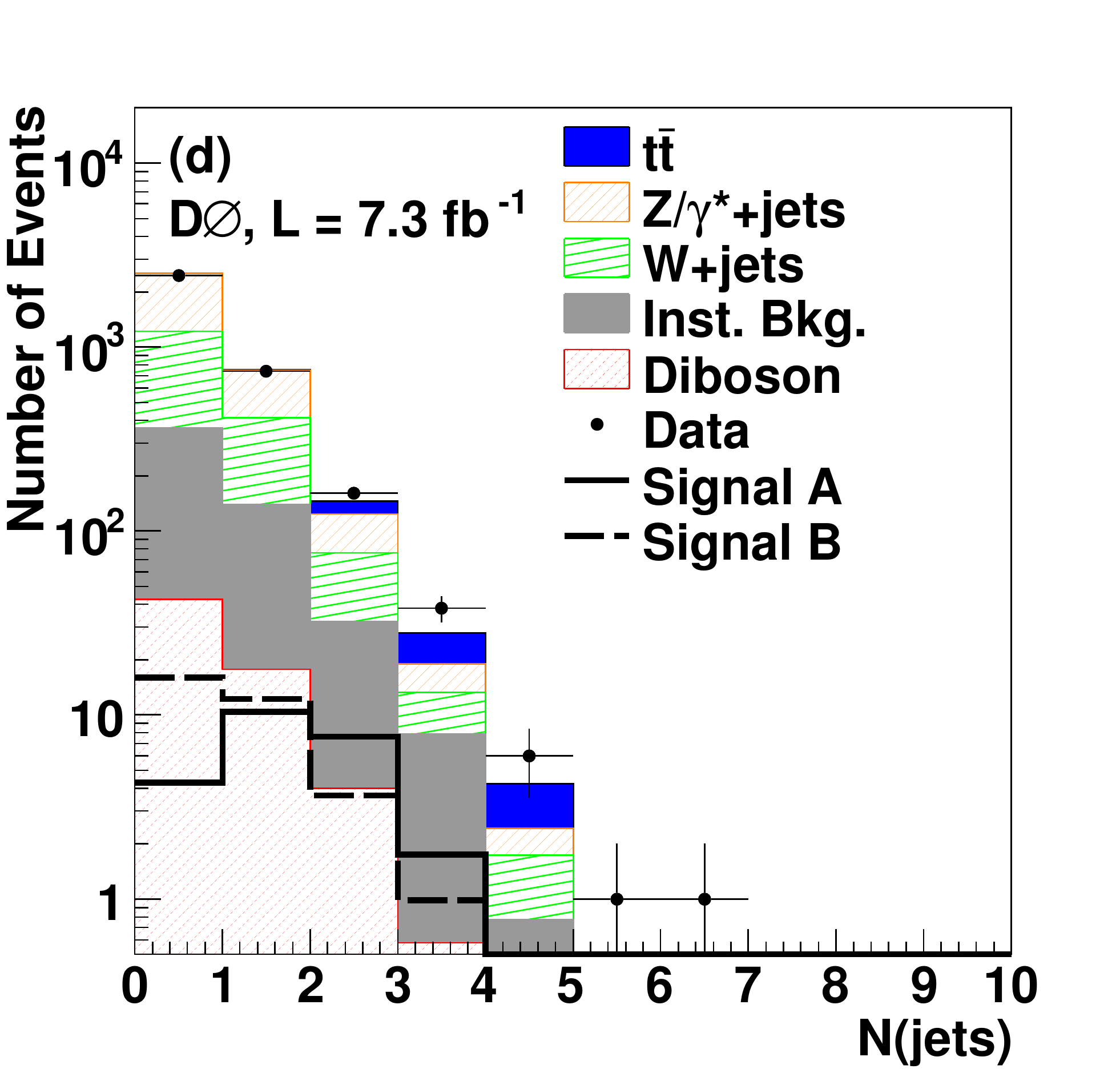}
\caption{\label{fig:presel} (Color online) Distributions of the number
  of jets after the preselection for (a) \mytau~type 1, (b)
  \mytau~type 2, (c) \mytau~type 3, and (d) their event sum. Signal A
and Signal B correspond to the wino scenario.}
\end{figure*}

%% file: data_mc_selection1.tex
\begin{table*}[!htpb]
 \caption{Numbers of events observed and expected from SM background
   processes and for the two signal samples A and B, after
   the final selection on \njets~$>$~0. The quoted uncertainties
   correspond to statistical sources.}\label{tab:selection1}
 \begin{tabular}{cD{,}{\,\pm\,}{-1}D{,}{\,\pm\,}{-1}D{,}{\,\pm\,}{-1}D{,}{\,\pm\,}{-1}D{,}{\,\pm\,}{-1}D{,}{\,\pm\,}{-1}D{,}{\,\pm\,}{-1}} 
\hline
\hline
\multicolumn{1}{c}{Process} & \multicolumn{1}{c}{\mytaua}& \multicolumn{1}{c}{\mytaub} & \multicolumn{1}{c}{\mytauc}  & \multicolumn{1}{c}{all \mytau} &  \multicolumn{1}{c}{\njets~=~1} & \multicolumn{1}{c}{\njets~=~2} & \multicolumn{1}{c}{\njets~$>$~2}\\
\hline
 \multicolumn{1}{c}{\ztautau} & \multicolumn{1}{c}{ 35.6}       & \multicolumn{1}{c}{ 226.2}  & \multicolumn{1}{c}{ 79.5}   & \multicolumn{1}{c}{ 341.3}  & \multicolumn{1}{c}{ 301.6}     &   \multicolumn{1}{c}{ 34.6} & \multicolumn{1}{c}{ 5.0} \\  
 \multicolumn{1}{c}{\zmumu} & \multicolumn{1}{c}{ 8.0}         & \multicolumn{1}{c}{ 20.7}   & \multicolumn{1}{c}{ 10.4}    & \multicolumn{1}{c}{ 39.1}   & \multicolumn{1}{c}{ 33.5}       &  \multicolumn{1}{c}{ 4.7}    & \multicolumn{1}{c}{ 0.9} \\
 \multicolumn{1}{c}{diboson } & \multicolumn{1}{c}{ 2.3}         & \multicolumn{1}{c}{ 12.5}   & \multicolumn{1}{c}{ 6.9}      & \multicolumn{1}{c}{ 21.7}   & \multicolumn{1}{c}{ 17.7}       & \multicolumn{1}{c}{ 3.3}     & \multicolumn{1}{c}{ 0.7} \\ 
 \multicolumn{1}{c}{\ttbar} & \multicolumn{1}{c}{ 3.0}             & \multicolumn{1}{c}{ 25.6}   & \multicolumn{1}{c}{ 10.1}    & \multicolumn{1}{c}{ 38.7}  & \multicolumn{1}{c}{ 8.1}       & \multicolumn{1}{c}{ 20.0}   & \multicolumn{1}{c}{ 10.6}\\
 \multicolumn{1}{c}{\wjets} & \multicolumn{1}{c}{ 30.4}           & \multicolumn{1}{c}{ 166.5}  & \multicolumn{1}{c}{ 117.5} & \multicolumn{1}{c}{ 314.4} & \multicolumn{1}{c}{ 301.6}       & \multicolumn{1}{c}{ 36.7}   & \multicolumn{1}{c}{ 6.0} \\
 \multicolumn{1}{c}{Instrumental} & \multicolumn{1}{c}{ 20.6} &
 \multicolumn{1}{c}{ 55.0}  & \multicolumn{1}{c}{ 74.1}    &
 \multicolumn{1}{c}{ 149.8} & \multicolumn{1}{c}{ 122.4}       & \multicolumn{1}{c}{ 20.1}   & \multicolumn{1}{c}{ 7.3} \\
\hline
 \multicolumn{1}{c}{Background total} & 99.9,2.3 & 506.6,5.4 & 298.5,4.9 & 905.1,9.6 & 755.1,8.4 & 119.4,2.0 & 30.6,0.8 \\
 \multicolumn{1}{c}{Data} & \multicolumn{1}{c}{90} & \multicolumn{1}{c}{532} & \multicolumn{1}{c}{271} & \multicolumn{1}{c}{893} & \multicolumn{1}{c}{738} & \multicolumn{1}{c}{116} & \multicolumn{1}{c}{39}\\
\hline
\multicolumn{1}{c}{Wino scenario} & & & & \\
 \multicolumn{1}{c}{Signal A} & 1.9,0.3  & 13.9,0.7          & 3.7,0.4          & 19.5,0.9        & 10.4,0.6       & 7.1,0.5        & 2.0,0.3\\
 \multicolumn{1}{c}{Signal B} &  2.0,0.9    & 10.6,2.1       & 3.6,1.2           & 16.3,2.6        & 12.2,2.3       & 3.1,1.1        & 1.0,0.6\\
\hline
\multicolumn{1}{c}{Higgsino scenario} & & & & \\
\multicolumn{1}{c}{Signal A} & 2.3,0.3  & 16.2,0.8      & 4.2,0.4         &  22.7,1.0     & 11.8,0.7              & 8.6,0.6        & 2.3,0.3\\
 \multicolumn{1}{c}{Signal B} &  1.8,0.8  & 9.9,2.0         & 3.1,1.1
 & 14.8,2.4      &10.9,2.0               & 2.9,1.0        & 1.0,0.6\\
\hline
\hline
  \end{tabular}
 \end{table*}

%% file: bdtvar.tex
\begin{table*}[!htpb]
 \caption{Listing of the five most sensitive input variables used for each
   BDT training and testing. The significance of \Met, Sig(\Met), is
   defined as the likelihood that the \Met~in an event is consistent
   with a fluctuation of the resolution on the \pt~measurements
   on the selected leptons and jets. $S_T$ is the sum of the lepton
   \pt~and of the \Met. The transverse mass $M_T$ is defined as
   \mbox{$M_T(A,B)=\sqrt{2p_T^Ap_T^B(1-\cos\Delta\phi(A,B))}$}. $H_T$
   is equal to the scalar sum of the $E_T$ of the jets. }\label{tab:bdtvars}
 \begin{tabular}{cccc|ccc}
\hline
\hline
 & \multicolumn{3}{c|}{Wino scenario} & \multicolumn{3}{c}{Higgsino scenario} \\
 & \njets = 1 & \njets = 2    & \njets $\geq$ 3  & \njets = 1 & \njets = 2 & \njets $\geq$ 3  \\
\hline 
               &  $S_T$                       & \deltaphiminjetmet                 &  \deltaphiminjetmet  & \deltaphiminjetmet &  \massmutau         & \deltaphiminjetmet\\

               & \etaleadjet                & \etaleadjet                               &  Sig(\Met)                 & \etaleadjet              &   \etaleadjet            & $H_T$\\ 
 
Signal A  &  \deltarmaxmujet      & \deltarmaxtaujet                     &   \deltarmaxtaujet     & \deltarmaxtaujet    & \deltarmaxmujet     & \deltarmaxtaujet\\

               & $M_T(\mu,\Met)$      &  $\Delta \phi (\mu\tau,\Met)$& \ettau                       & Sig(\Met)                 & \deltaphimutaumet   & \massmutau \\

                &  \massleptjet            &  \deltarmaxmujet                  & \pt(leading jet)           & \massleptjet           & $M_T(\tau,\Met)$     & \massmujet\\

\hline

               & Sig(\Met)              & \deltaphiminjetmet   & \deltaphiminjetmet    &    $\Delta \eta(\mu,\tau)$      & \deltaphiminjetmet    & \deltaphimutau \\

               & $\eta (\tau)$        & \etaleadjet                & Sig(\Met)                    & $S_T$                                       & \etaleadjet                  & \deltaphileadjetmet\\

Signal B  & $S_T$                    & \deltarmintaujet       &  \deltaphimutaumet    & $\eta(\tau)$                           & \deltarmintaujet           & \deltarmaxtaujet \\

              &  \deltaphimutau   & \deltaphinleadjetmet& \deltaphileadjetmet     &  \deltaphimutau                     & \deltaphileadjetmet    & \deltaphimumet \\

               & \massleptjet       & \deltarminmujet        &  \massmujet               & \massleptjet                          &   \deltaphimutaumet    & $\Delta \phi (\tau,\Met)$ \\
\hline
\hline

  \end{tabular}
 \end{table*}


%% file: plots_bdt.tex
\begin{figure*}
\hspace*{-0.5cm}
\includegraphics[scale=0.3]{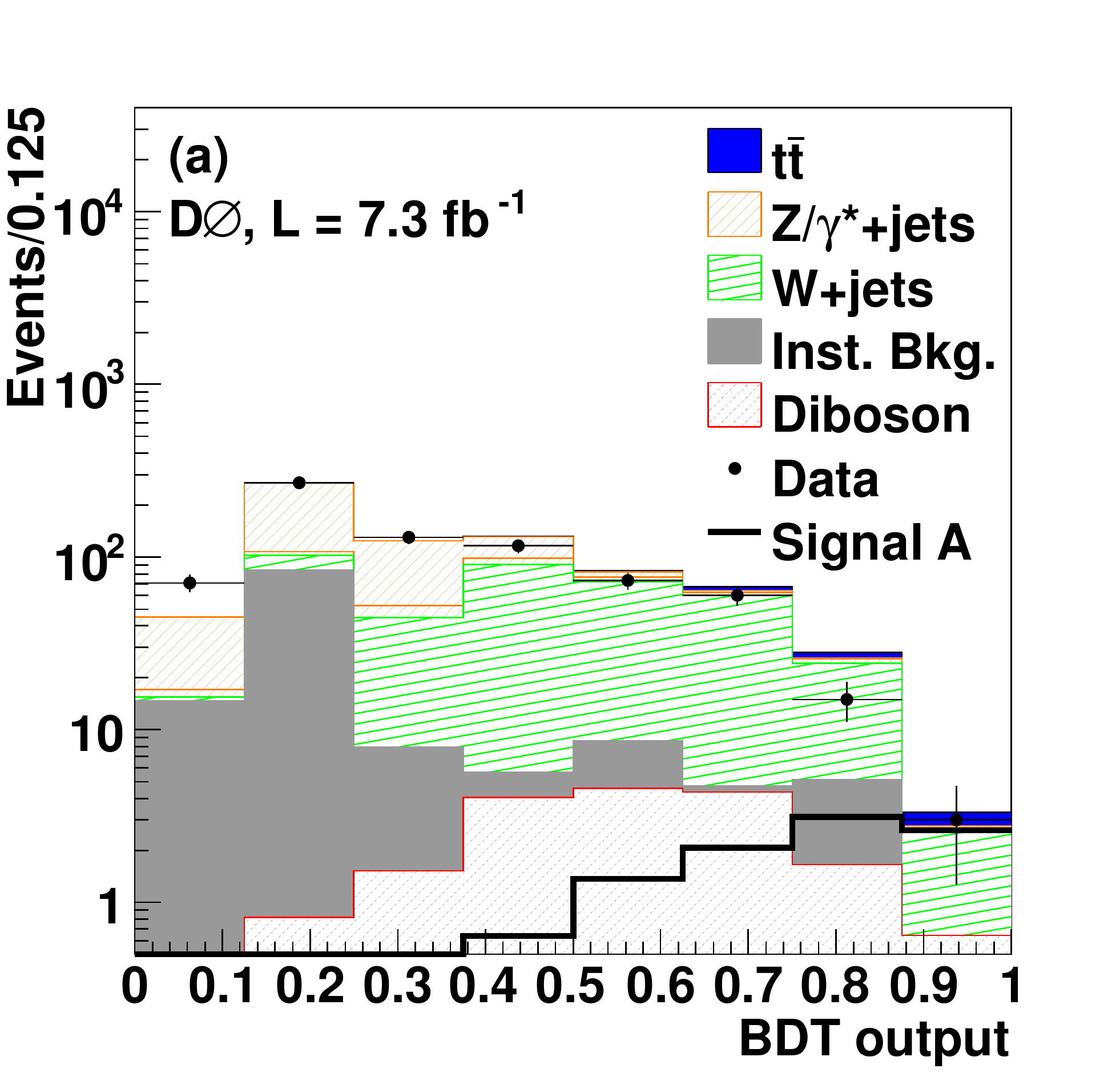}
\includegraphics[scale=0.3]{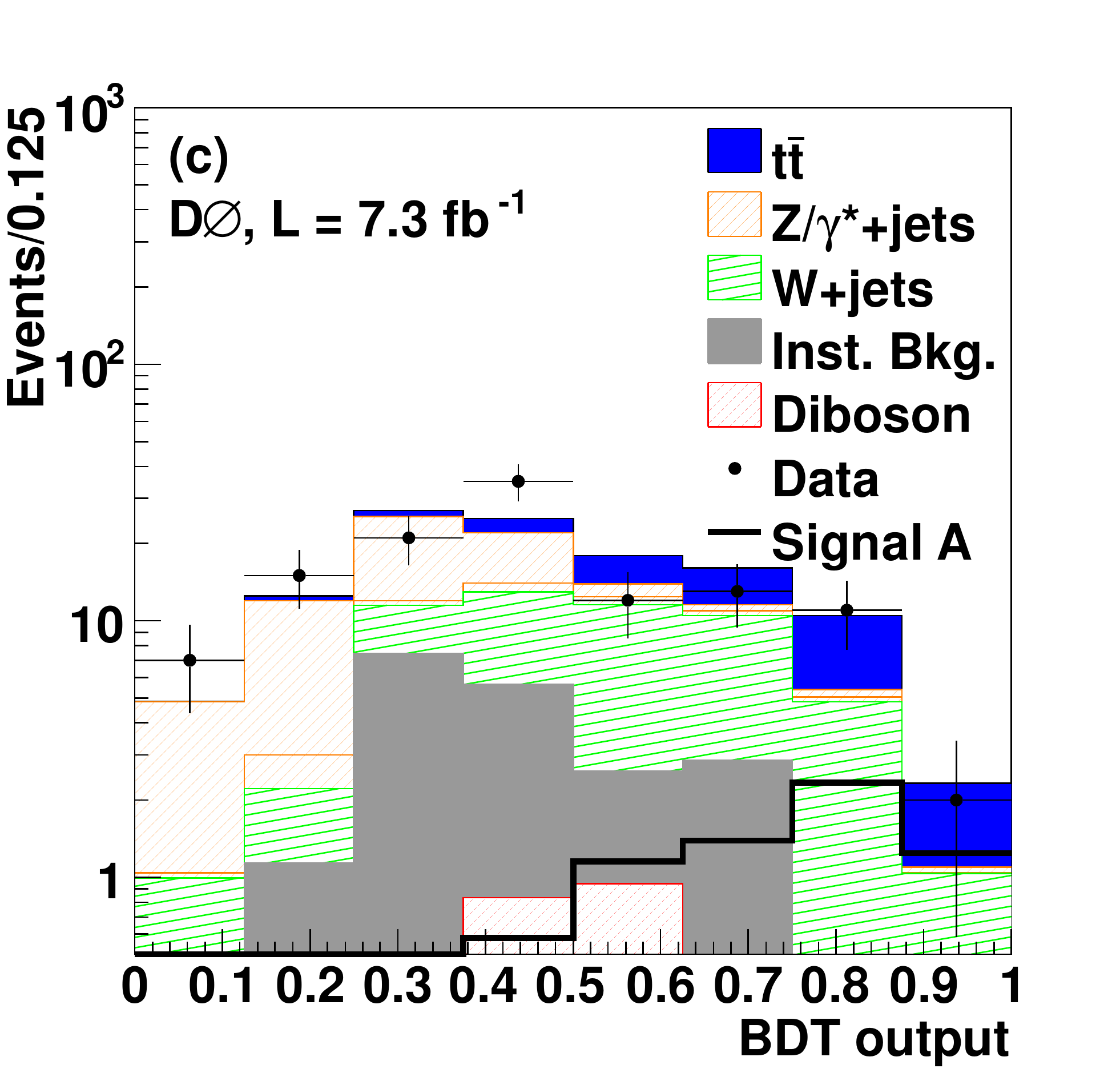}
\includegraphics[scale=0.3]{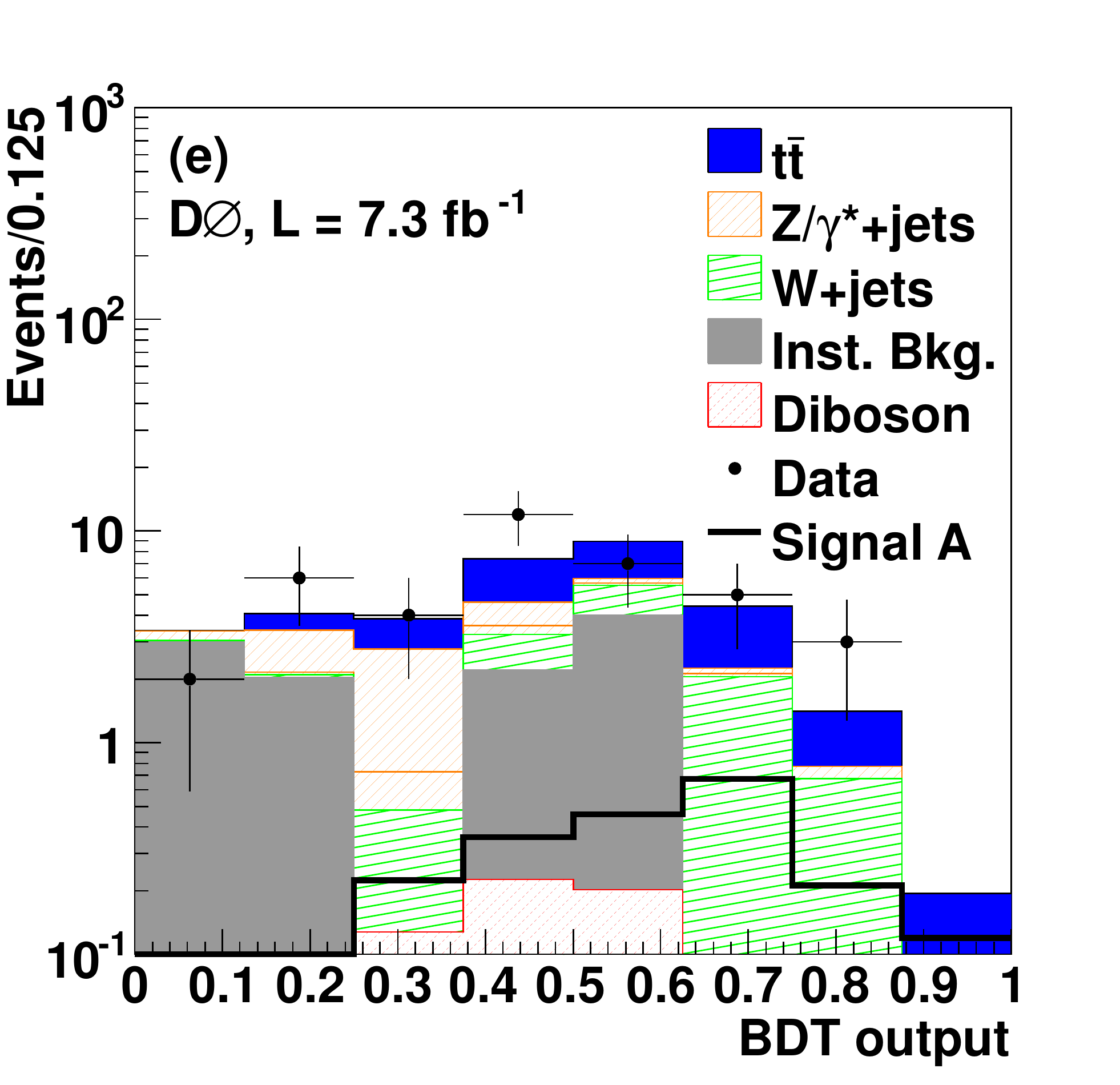}
\hspace*{-0.5cm}
\includegraphics[scale=0.3]{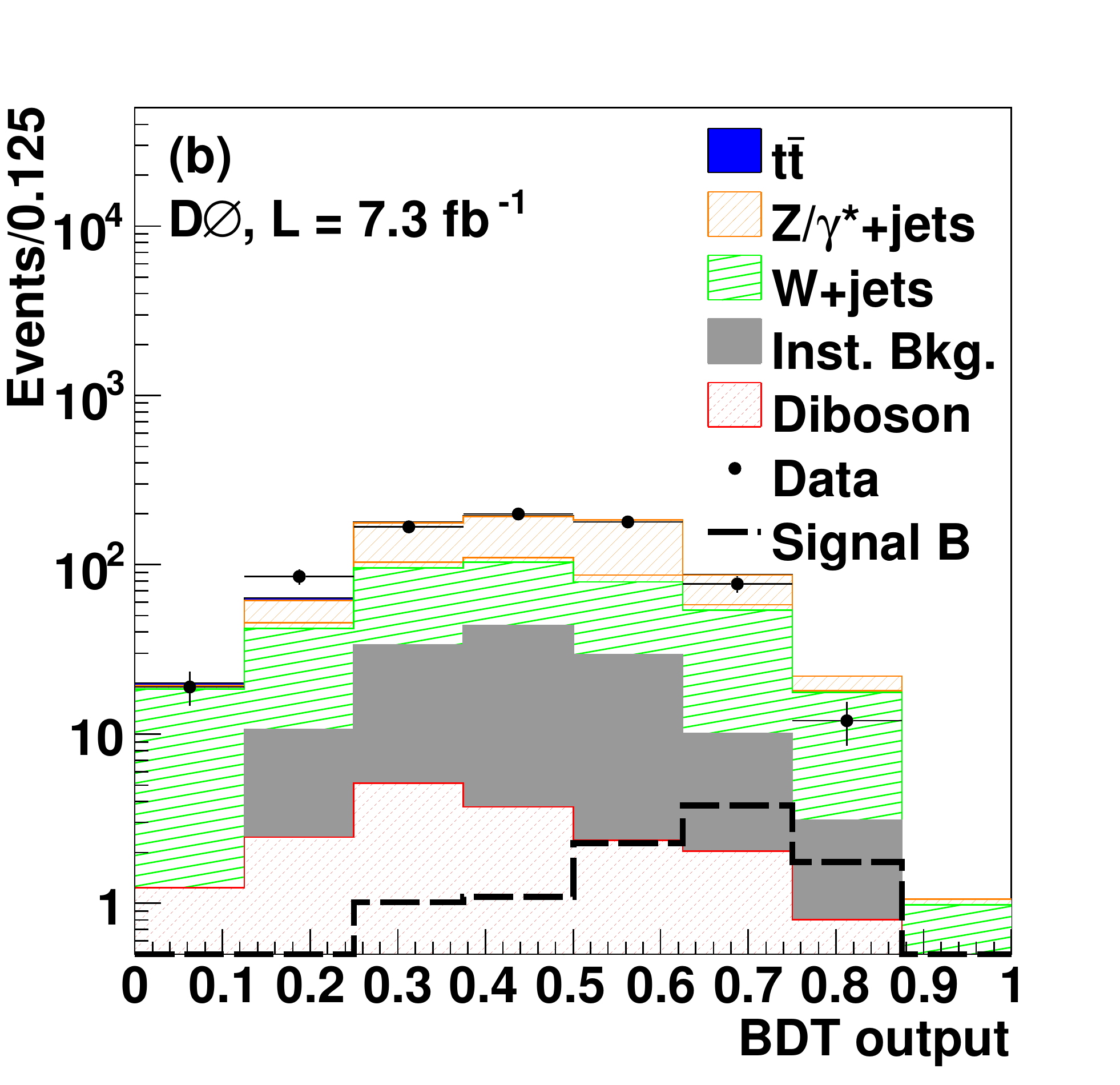}
\includegraphics[scale=0.3]{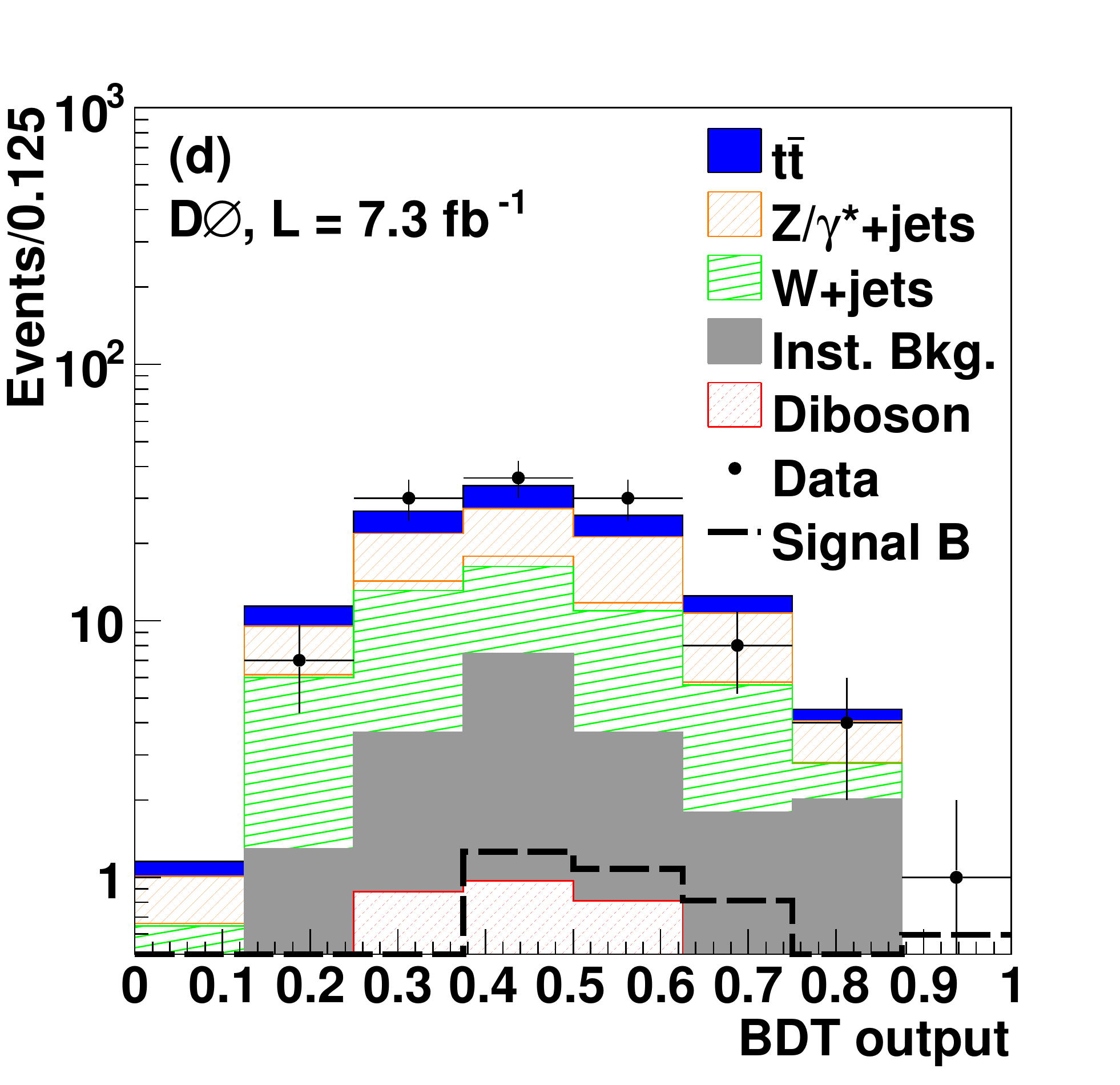}
\includegraphics[scale=0.3]{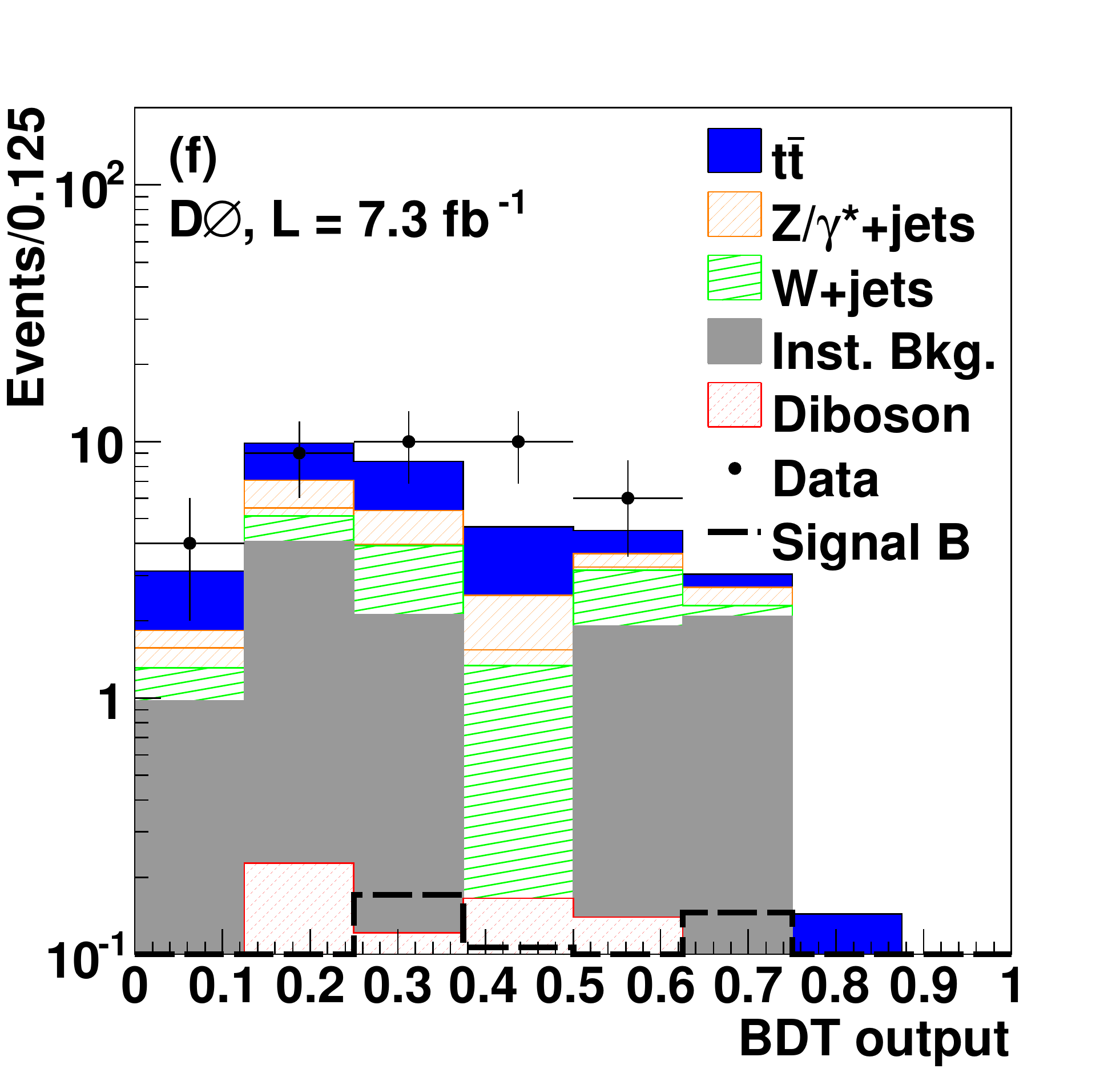}
\caption{\label{fig:bdts_wino} (Color online) Distributions of the BDT
  output discriminants, in the wino scenario,  for the sample with \njets
  = 1, (a) for Signal A, (b) for Signal B;  \njets = 2, (c) for Signal
  A, (d) for Signal B;  \njets $>$ 2,  (e) for Signal A, (f) for Signal B.}
\end{figure*}

\begin{figure*}
\hspace*{-0.5cm}
\includegraphics[scale=0.3]{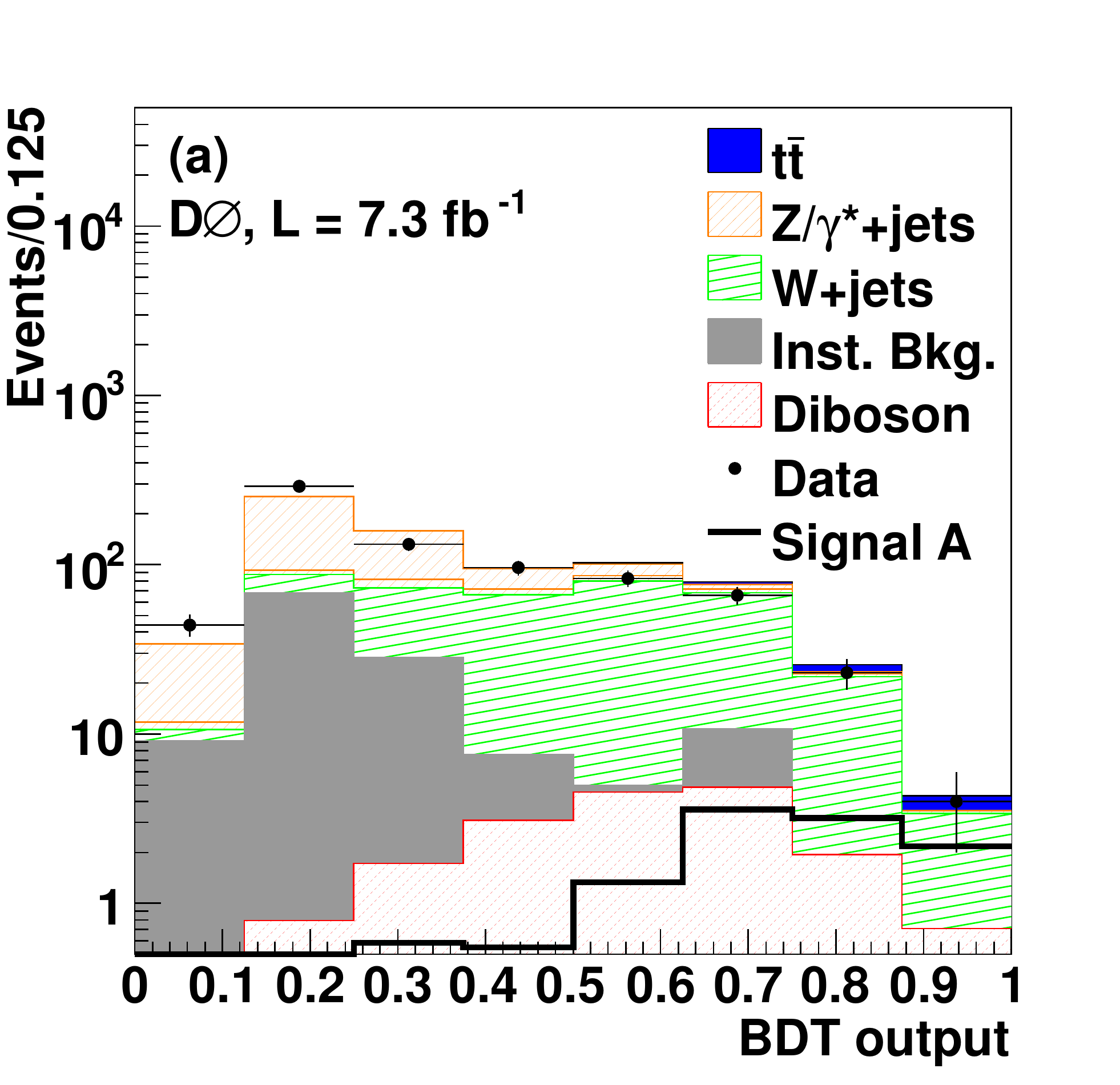}
\includegraphics[scale=0.3]{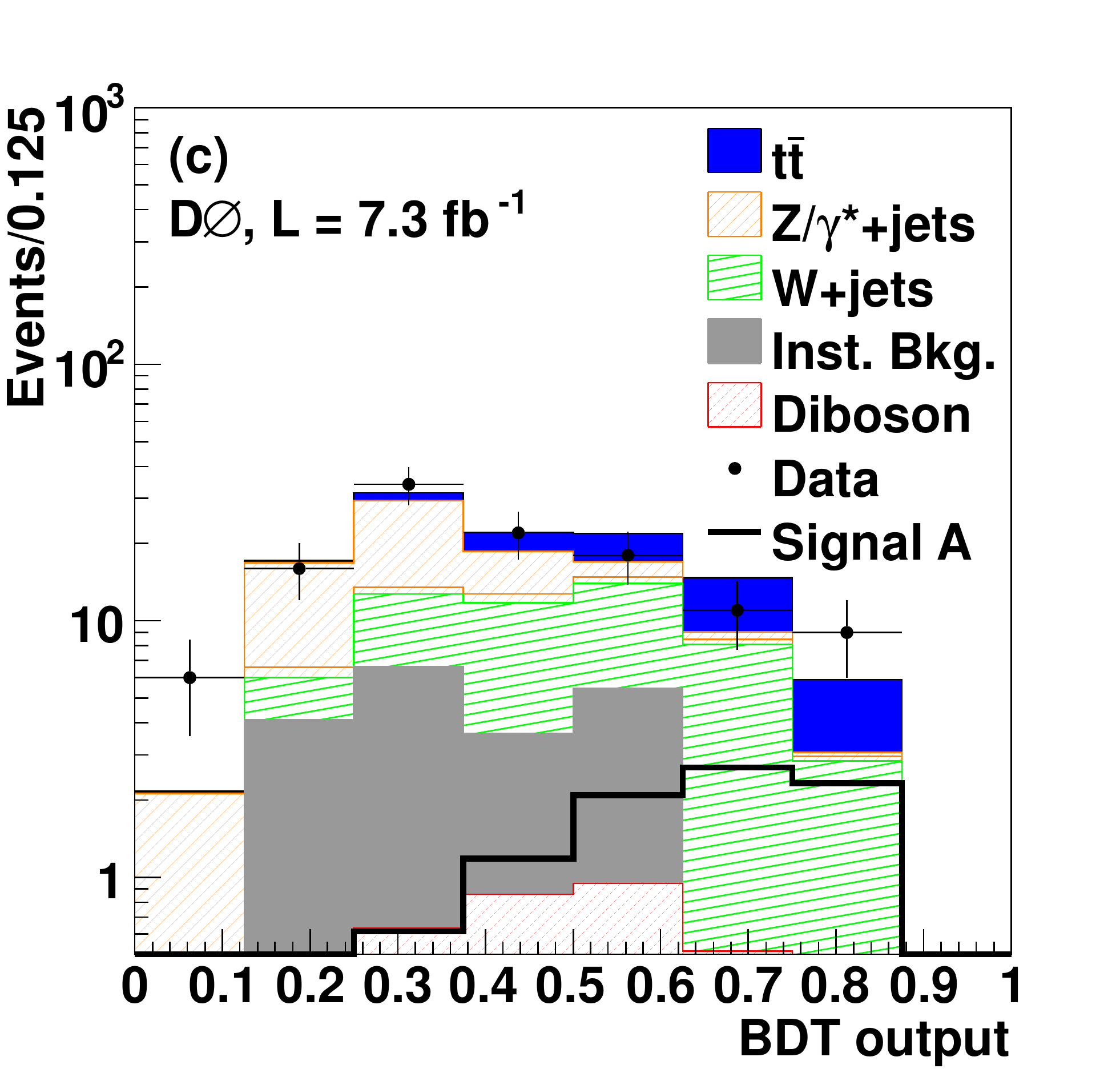}
\includegraphics[scale=0.3]{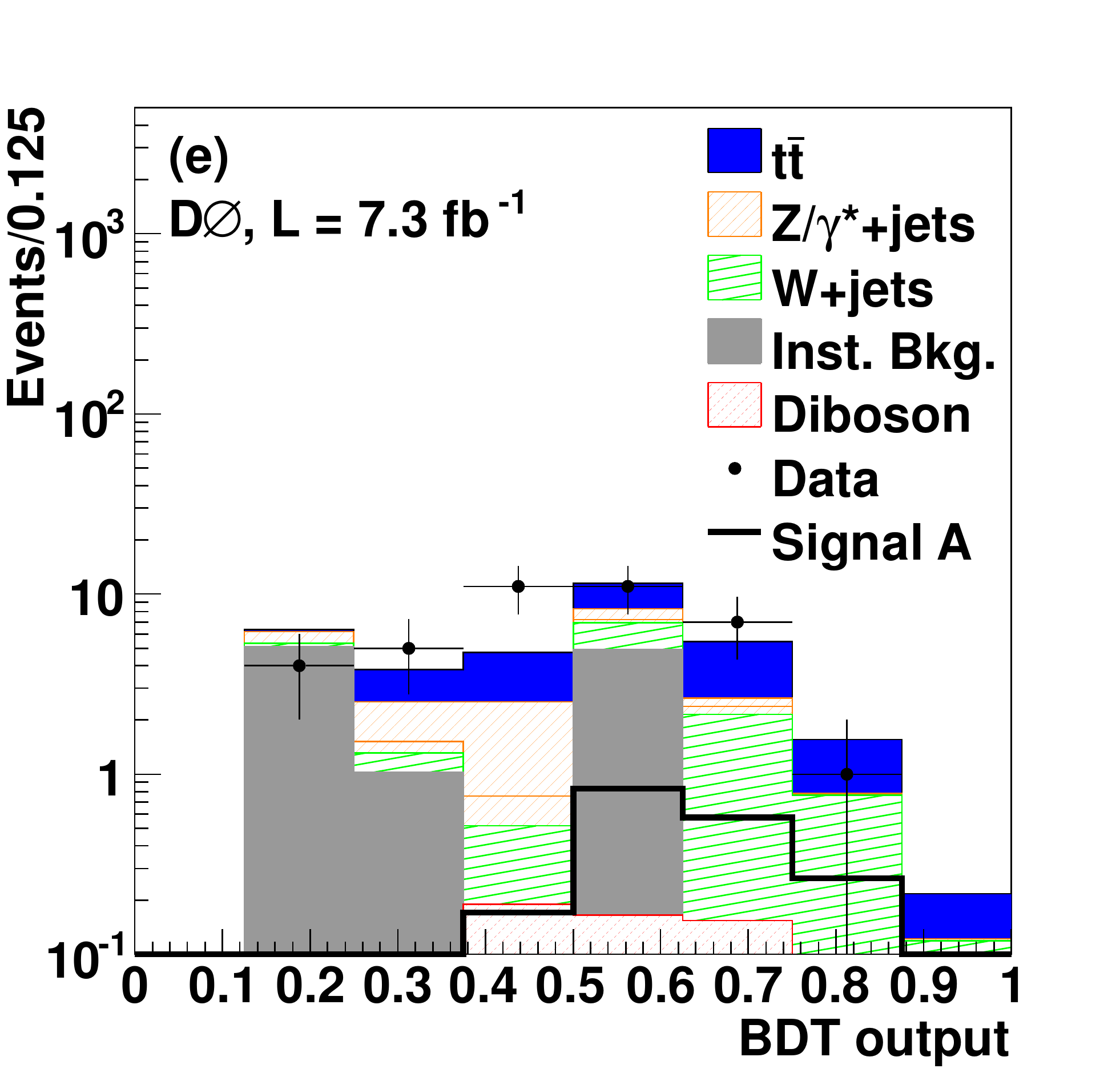}
\hspace*{-0.5cm}
\includegraphics[scale=0.3]{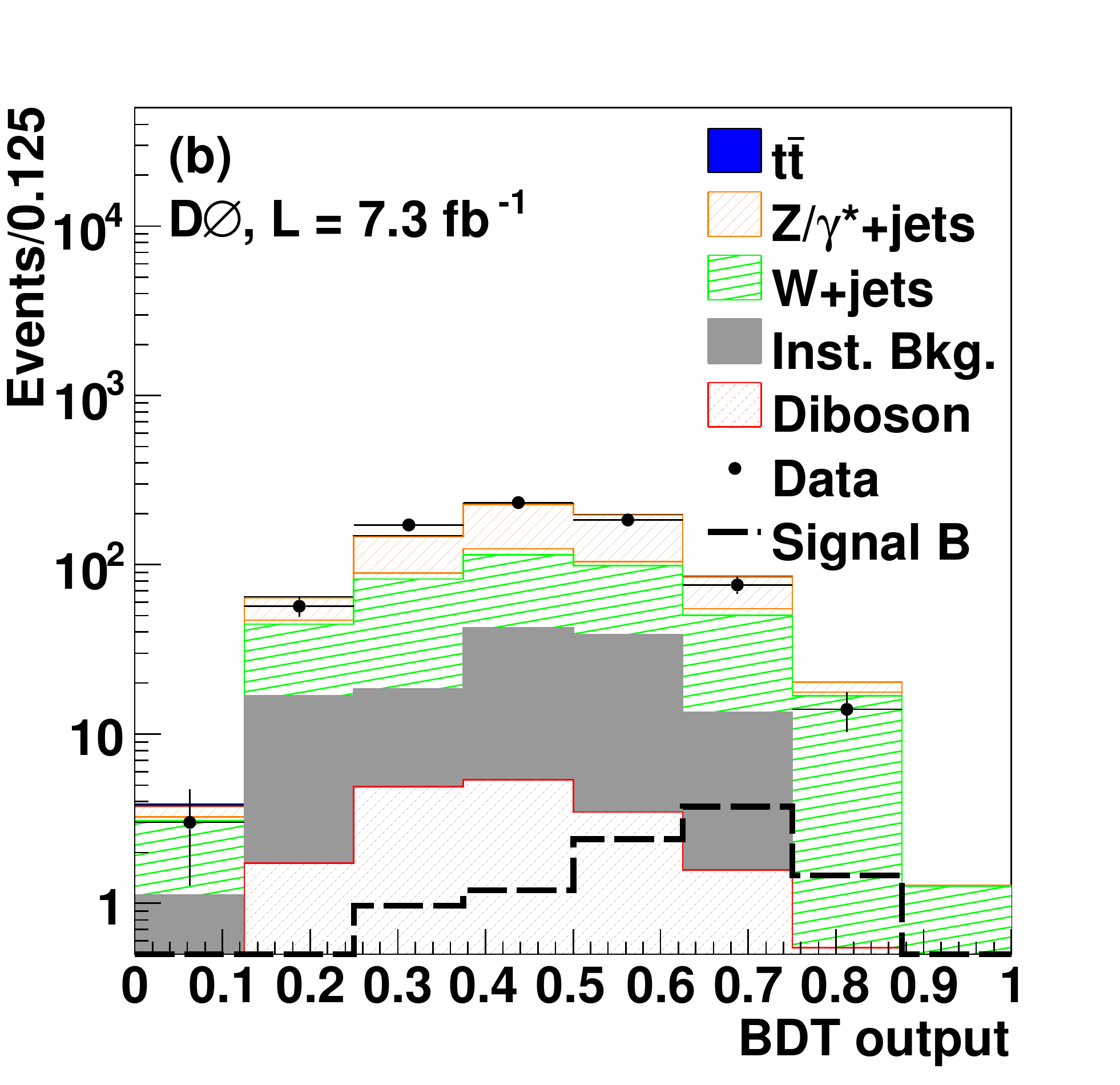}
\includegraphics[scale=0.3]{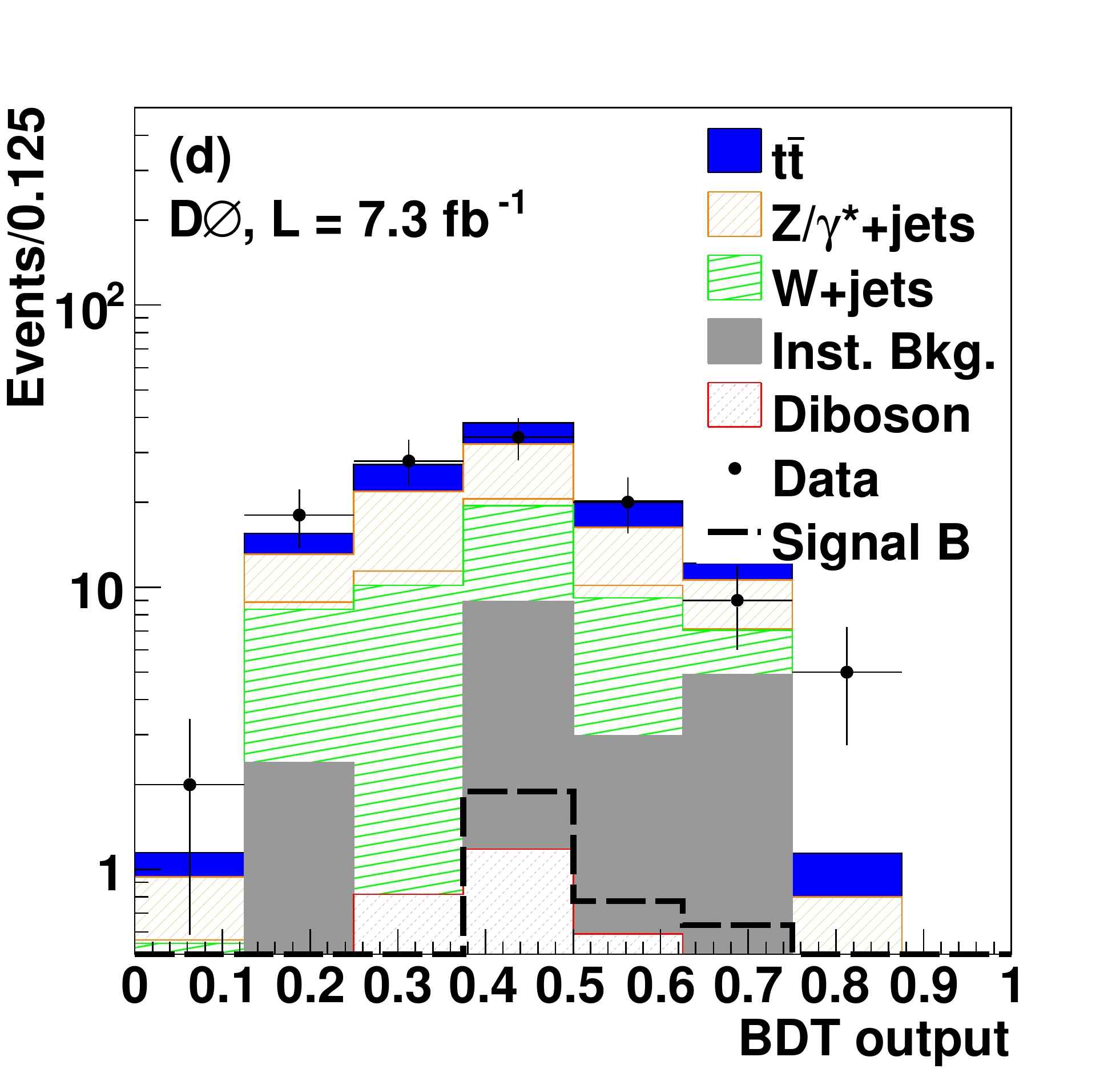}
\includegraphics[scale=0.3]{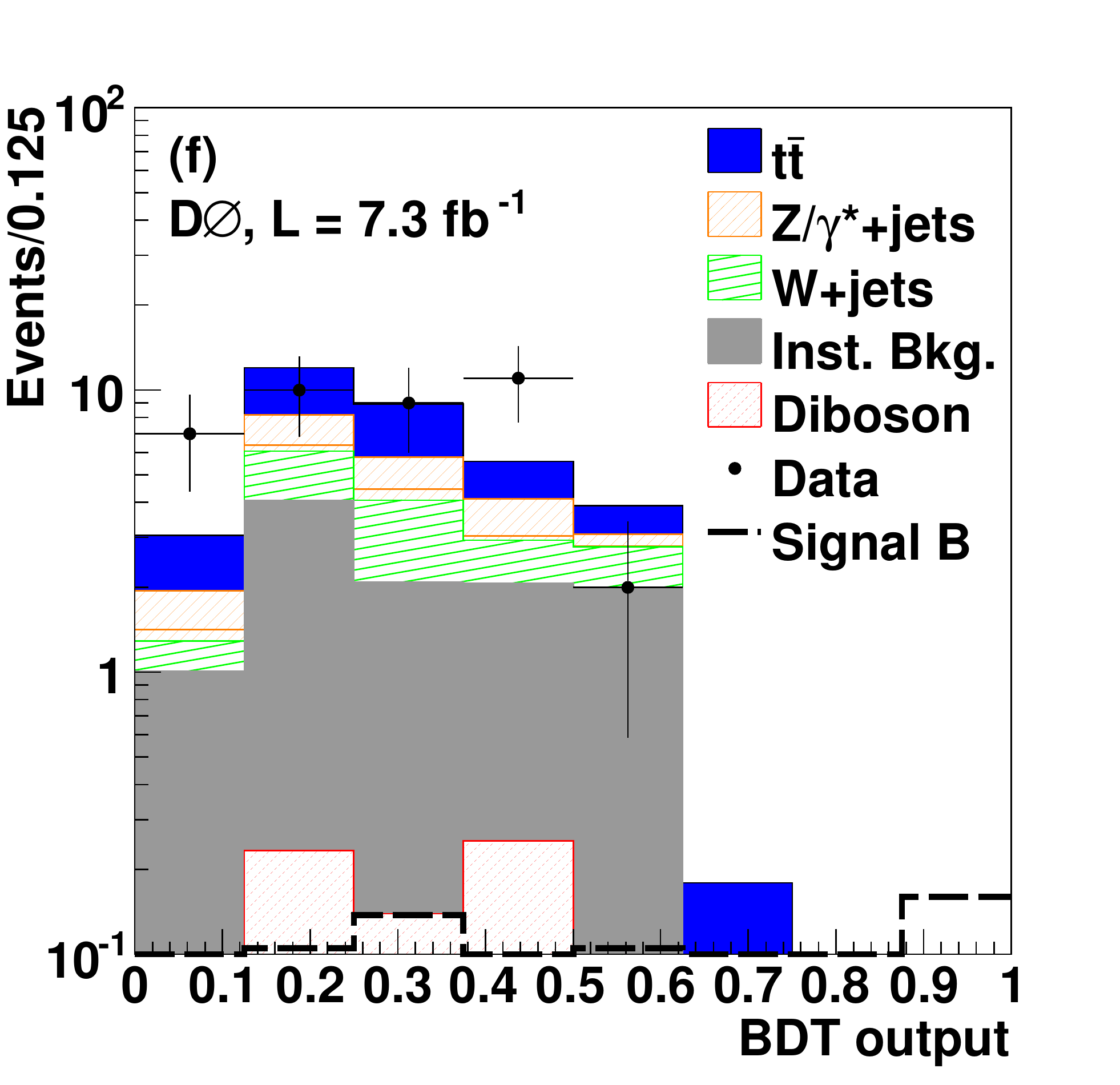}
\caption{\label{fig:bdts_higgsino} (Color online) Distributions of the BDT
  output discriminants, in the higgsino scenario, for the sample with \njets = 1,  (a) for Signal A, (b) for Signal B;  \njets = 2,  (c) for Signal A, (d) for Signal B;  \njets $>$ 2,  (e) for Signal A, (f) for Signal B.}
\end{figure*}

%% file: exclusion_nosigma.tex
\begin{figure}[!htpb]
\includegraphics[width=10.cm]{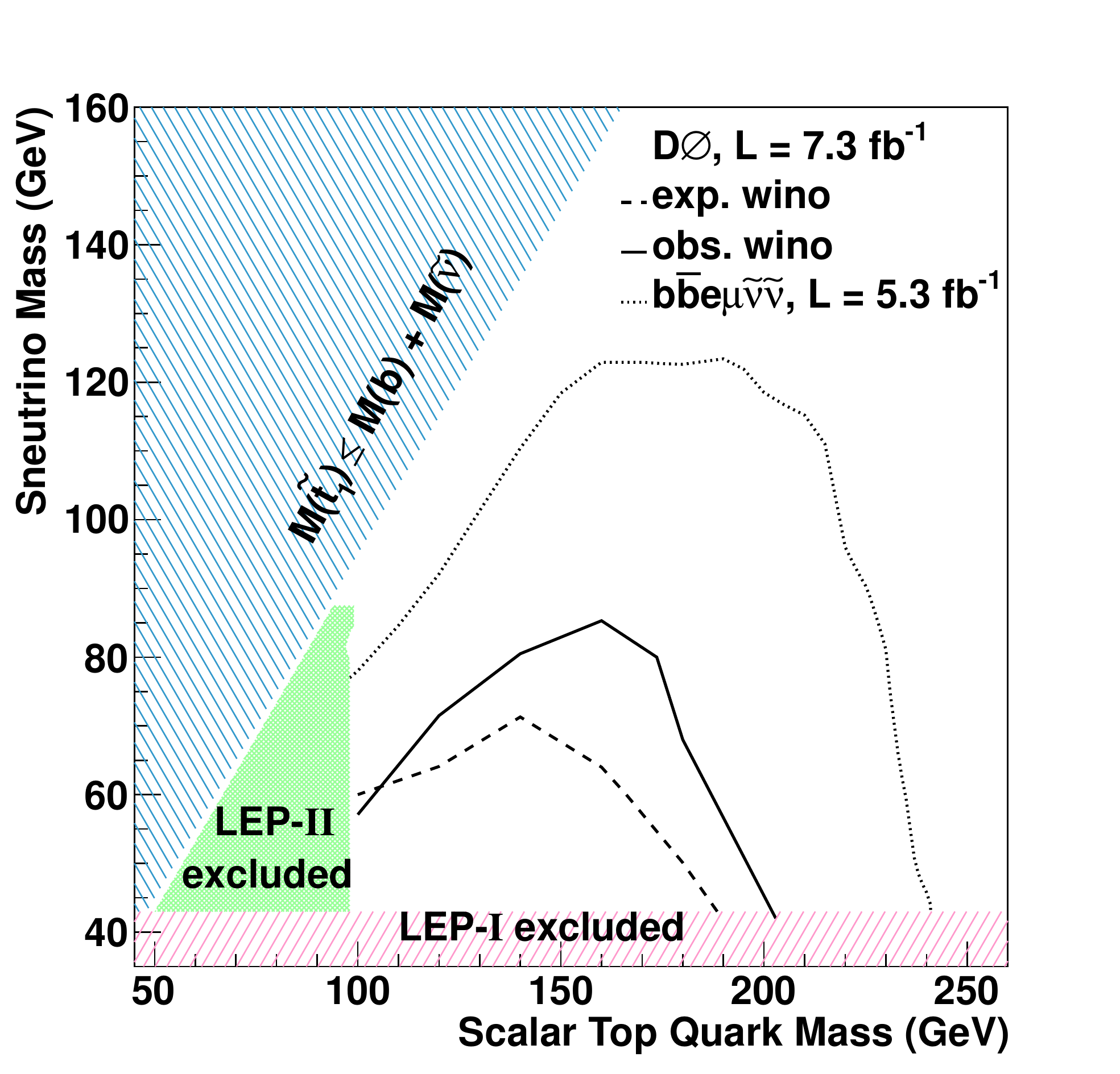}
\caption{(Color online) The 95\% CL contour of exclusion in the sneutrino versus scalar top quark mass plane obtained for the assumption $B$(\stopa~$\rightarrow$~\bmusneut) = $B$(\stopa~$\rightarrow$~\btausneut)  = 1/3 (wino scenario). Shaded areas represent the kinematically forbidden region and the LEP-I~\cite{stoplep1} and \mbox{LEP-II~\cite{lepstop}} exclusions. The dashed and continuous lines represent, respectively, the expected and observed 95\% CL exclusion limits for this analysis.
The region excluded by a recent \dzero~search \cite{dzerostop3} in the \stopa\astopa~$\rightarrow$~\bbemu~final state in the wino scenario is indicated by the dotted line.}\label{fig:limite_wino}
\end{figure}

\begin{figure}[!htpb]
\includegraphics[width=10.cm]{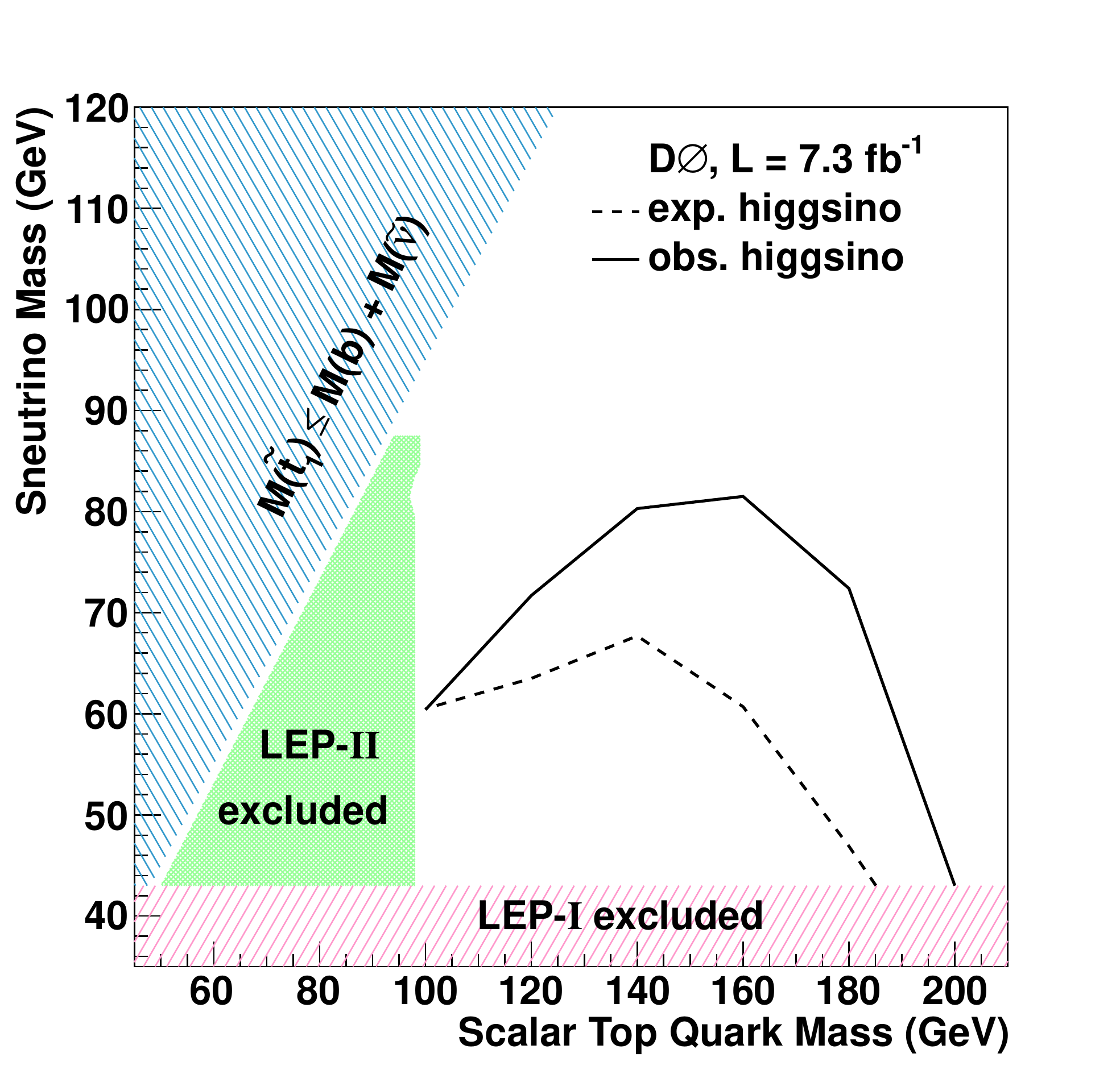}
\caption{(Color online) The 95\% CL contour of exclusion in the sneutrino versus scalar top quark mass plane obtained for the assumption $B$(\stopa~$\rightarrow$~\bmusneut) = 0.1 and  $B$(\stopa~$\rightarrow$~\btausneut)  = 0.8 (higgsino scenario). Shaded areas represent the kinematically forbidden region and the LEP-I~\cite{stoplep1} and \mbox{LEP-II~\cite{lepstop}} exclusions. The dashed and continuous lines represent, respectively, the expected and observed 95\% CL exclusion limits for this analysis.}\label{fig:limite_higgsino}
\end{figure}

%% file: acknowledgement.tex
%
We thank the staffs at Fermilab and collaborating institutions,
and acknowledge support from the
DOE and NSF (USA);
CEA and CNRS/IN2P3 (France);
FASI, Rosatom and RFBR (Russia);
CNPq, FAPERJ, FAPESP and FUNDUNESP (Brazil);
DAE and DST (India);
Colciencias (Colombia);
CONACyT (Mexico);
NRF (Korea);
CONICET and UBACyT (Argentina);
FOM (The Netherlands);
STFC and the Royal Society (United Kingdom);
MSMT and GACR (Czech Republic);
BMBF and DFG (Germany);
SFI (Ireland);
The Swedish Research Council (Sweden);
and
CAS and CNSF (China).

%% file: Paper_Stop_mu_tau.bbl
\begin{thebibliography}{99}
\bibitem{pseudo}
The pseudorapidity is defined as $\eta~=~-\ln[\tan(\theta/2)]$, with $\theta$ being the polar angle with respect to the proton beam direction, relative to the center of the detector.
\end{thebibliography}
